\documentclass[twocolumn,amssymb,amsmath,
nofootinbib,tightenlines,showpacs,floatfix,
superscriptaddress,aps,prb]{revtex4-1}
\usepackage{amsfonts,amsbsy,graphicx,bm,color,txfonts,dcolumn,times,bbm}
\usepackage[colorlinks=true,linkcolor=blue,pagecolor=blue,filecolor=blue,menucolor=yellow,urlcolor=blue,citecolor=blue,anchorcolor=blue]{hyperref}%
\begin{document}
\def\bar{\begin{eqnarray}}
\def\ear{\end{eqnarray}}
\def\beq{\begin{equation}}
\def\eeq{\end{equation}}
\newcommand{\degrees}{\ensuremath{^{\circ}}}
\newcommand{\bsigma}{\mbox{\boldmath$\sigma$}}
\newcommand{\bnabla}{\mbox{\boldmath$\nabla$}}
\newcommand{\identity}{\mathbbm{1}}
\newcommand{\ttensor}[1]{\overline{\overline{#1}}}

\title{Charge and Spin Currents in 
Ferromagnetic Josephson junctions}
\author{Klaus Halterman }
\email{klaus.halterman@navy.mil}
\affiliation{Michelson Lab, Physics
Division, Naval Air Warfare Center, China Lake, California 93555}
\author{Oriol T. Valls}
\email{otvalls@umn.edu}
\altaffiliation{Also at Minnesota Supercomputer Institute, University of Minnesota,
Minneapolis, Minnesota 55455}
\affiliation{School of Physics and Astronomy, University of Minnesota, 
Minneapolis, Minnesota 55455}
\author{Chien-Te Wu}
\altaffiliation{School of Physics and Astronomy, University of Minnesota,
Minneapolis, Minnesota 55455}
\affiliation{The James Franck Institute, The University of Chicago, Chicago,
Illinois, 60637}
\email{chientewu@uchicago.edu}
\date{\today}


\begin{abstract} 
We 
determine, using a self consistent method,  the 
charge and spin currents in 
ballistic  
Josephson junctions consisting of  several 
ferromagnetic ($F$) layers sandwiched between superconducting
($S$) electrodes 
($SFS$-type junctions). When there are two $F$ layers,
we also consider the experimentally relevant 
configuration where
a normal ($N$) nonmagnetic spacer layer separates them.
We study the
current-phase relationships as functions  
of geometrical parameters that are accessible experimentally 
including particularly the angles that characterize the
relative orientation of the magnetization in the $F$ layers.
Our self-consistent method ensures that the proper charge conservation laws
are satisfied, and that important proximity effects are fully and 
properly accounted for.
We find that as we vary
the phase difference $\Delta\varphi$ between the two outer 
$S$ electrodes,
multiple harmonics in the current phase relations emerge, 
the extent of which depends
on the interface scattering strength and on the
relative $F$ layer widths and magnetization orientations. 
By manipulating the relative $F$ layer magnetization orientations, 
we find that the charge supercurrent can reverse directions or 
vanish altogether.
These findings are discussed in the context of the 
generation and long-range nature of 
triplet pair correlation within these structures.
We  also investigate the 
spin currents and 
associated spin transfer torques throughout 
the entire junction regions.
For noncollinear relative magnetizations, 
the non-conserved spin currents in a given $F$ region gives rise to
net torques that can switch directions at  particular  magnetic configurations
or  $\Delta\varphi$ values. The 
details of
the spin current behavior 
are shown to depend strongly on the degree of magnetic inhomogeneity 
in the system,
including the number of $F$ layers and the relative widths 
of the $F$ and $N$ layers.
\end{abstract}
\maketitle

\section{Introduction}
\label{introduction}

When a phase difference, $\Delta \varphi$, 
exists between two superconductor ($S$)
electrodes separated by a non-superconducting 
material in a Josephson junction,
the corresponding charge supercurrent
is directly controllable via $\Delta \varphi$.
Motivated by the interplay between ferromagnetism and superconductivity,
researchers are also interested in
the dc Josephson effect in superconducting junctions that
contain a central ferromagnet ($F$) region, which in turn can give
rise to an additional spin degree of freedom. More specifically, 
this kind of Josephson effect 
provides a venue for
the study of spin 
currents  that
can be manipulated in cryogenic spintronic systems.~\cite{eschrig,hikino,grein}
Besides numerous practical applications\cite{golubov} 
involving these $SFS$-based Josephson junctions, 
it is found that novel and 
interesting phenomena can arise. For example, the realization of a $\pi$
state,~\cite{ryazanov,golubov,ah,hv4,hv5}
where the ground state of the system corresponds to $\Delta\phi=\pi$
across the junction.
Moreover,
if the ferromagnet region consists of 
at least two $F$ layers that each have a uniform magnetizations (e.g., a $SFFS$ structure),
manipulation of the angle  between the magnetization vectors
can serve to generate long range triplet 
supercurrents~\cite{bergeret,trifunovic,pajovic,Shomali,baker,hbvprl} 
in addition to the ordinary singlet 
ones. 
Additional control of the magnetic state can also
occur from the spatially varying spin current  within the $F$ layers
of the junction, 
causing mutual  torques to act on
their respective magnetic moments.
Therefore, 
$SFS$-based junctions
that contains
multiple $F$ layers,
present many  opportunities 
for controlling
the charge and spin currents, 
and their influence on the magnetization 
in terms of the torque they produce.

Although  interest
in the study of $F/S$ multilayer structures
has recently  increased considerably, work on  
$SFS$ Josephson junctions actually started long
ago. The Josephson and critical current oscillations
were found to occur as a function of 
the ferromagnet exchange field\cite{buz,buz2} 
and the thickness of the magnet\cite{buz2}.
An essential principle behind many
of these important phenomena 
is the damped oscillatory nature
of the singlet Cooper pairs in the ferromagnetic regions, and 
the associated phase shift in the superconducting order
parameter. Due to the intrinsic exchange field in
the ferromagnet regions, electrons of Cooper pairs
with spin-up (down) decrease (increase)
their kinetic energy and the  Cooper pairs acquire  a 
nonzero center-of-mass momentum. It further
leads to an oscillatory order parameter in 
the $F$ regions~\cite{demler}. 
Owing to this oscillatory nature,
not only the Josephson critical current but also
the superconducting critical temperatures oscillate
as a function of exchange field and thickness of magnets.
Josephson junctions of the $\pi$ type can also be realized by 
using this principle by adjusting either the exchange field, the
magnet width, or both~\cite{hv5,robinson2,linder2}.  
The proximity effects between the $S$ and 
$F$ regions thus  give rise
to phenomena~\cite{buzdin1,demler,hv1,klaus}
that subsequently play  crucial roles in
the charge and spin currents
that may be manipulated in low temperature nanoscale devices, including  
nonvolatile
memory elements, where
the dissipationless nature of
the supercurrent flow offers 
reduced energy loss and Joule heating.


In equilibrium,
singlet Cooper pairs
carry no net spin, hence any 
spin current in the system 
either flows only within the 
ferromagnets 
due to their exchange interaction, or it flows by means
of induced 
equal-spin triplet correlations, where the Cooper pairs 
have a net spin of $m=\pm1$ on the spin quantization axis
and they can reside in both $S$ and $F$ regions.
The generation of long-range triplet
proximity effects in superconducting heterostructures with magnetic
inhomogeneity has been both theoretically predicted and
experimentally confirmed: by introducing magnetic inhomogeneity, 
e.g., inclusion of an  additional magnet with  misaligned 
exchange field, the Hamiltonian no longer commutes with the
total spin operator and equal-spin triplet correlations can then be
induced.~\cite{bergeret2,hvb08,hbvprl} 
Due to the imbalance between majority
and minority spins in a ferromagnet, conventional singlet 
superconducting correlations do not survive for long once
inside the magnetic region. However, 
Cooper pairs with electrons
that carry the same spin are not subject to paramagnetic
pair-breaking and can in principle propagate for
large distances inside the ferromagnet, 
limited only by coherence
breaking
processes.~\cite{bergeret,bergeret2,hvb08,hbvprl}
Such equal-spin triplet correlations thus
play an important role in  
Josephson junctions with inhomogeneous ferromagnets.~\cite{bergeret,alidoust_spn}
Indeed, it has been reported  experimentally that with the presence 
of magnetic inhomogeneity, 
the Josephson critical current decays much more slowly with  increases in
the F layer thicknesses,~\cite{robinson,khaire} as compared to junctions
with homogeneous magnetization. 
One of the simplest ways to introduce magnetic inhomogeneity
in a Josephson junction,
is through the insertion of bilayer or trilayer of uniformly magnetized ferromagnets. 
Experimentally,
this has the advantages of reproducibility 
and easy  manipulation of the relative  
exchange field orientation. 
These structures also provide direct evidence of 
triplet correlations.~\cite{ilya,wvh14,volkov,birge}
Recently, long-range coherent transport of triplet pairs 
was studied in double magnet $S F_1 F_2 S$ junctions,~\cite{richard,hikino}
further 
demonstrating that it is not always necessary to have a trilayer\cite{buzdin} ferromagnet structure 
of misaligned ferromagnets to generate equal-spin triplet components to the supercurrent.
This finding was in consistent  with the long-range phenomena found in 
a similar structure~\cite{trifunovic} 
with asymmetric widths,~\cite{trifunovic} 
and 
orthogonal 
exchange fields.
The behavior of the triplet amplitude is often 
anticorrelated\cite{wvh14} to that of the critical
temperature and associated singlet correlations, suggesting
singlet to triplet conversion.

The oscillatory and long-range pair correlations
also lead to new behaviors in the 
current-phase relations (CPRs) in $SFS$ juncitons~\cite{pajovic,golubov,gold}
with a nontrivial magnetic structure. 
The CPR can in general  contain not only the first harmonic 
but also higher order harmonics, i.e. 
$I(\Delta\varphi)\approx I_1\sin(\Delta\varphi)+I_2\sin(2\Delta\varphi)$.
The appearance of additional harmonics
in the current phase relation has also been discussed in the diffusive
and clean regimes for  ferromagnetic Josephson
junction structures.~\cite{buzdin1,2nd_hrmnc_3,zareyan,alidoust_spn} 
In  conventional Josephson junctions without
any magnetic interaction, the magnitude of $I_2$ is much smaller than 
that of $I_1$. However, in 
$SFS$ Josephson junctions, near a $0$ to $\pi$ phase 
transition, the roles of the first harmonic and the second harmonic
can be reversed, and the CPR can be largely
dominated by the second harmonic~\cite{2nd_hrmnc_3}. In this regime,
both $\Delta\varphi=0$ and $\Delta\varphi=\pi$ states can 
be stable or metastable, and they can coexist. 
The physical origins of the second and higher order harmonics
are believed to lie 
in the long ranged triplet component of the supercurrent.
In this regard, within the vicinity of  the $0$-$\pi$ transitions,
the triplet correlations can be tuned accordingly.
Since 
 the first harmonic is 
suppressed due to the supercurrent flow reversing direction,
the higher order harmonics are revealed at
the 0-$\pi$ transition point.
The influence of interface scattering on
the higher order harmonics  were
 investigated  in the quasiclassical clean limit\cite{2nd_hrmnc_1} and
experimentally detected.~\cite{2nd_hrmnc_2} 
 The
measured  supercurrent at the
0-$\pi$ transition point~\cite{ryazanov} was
attributed to the presence of higher harmonics.
Subsequent work with ferromagnetic Josephson junctions
demonstrated that the higher harmonics can naturally arise
when varying the location of domain walls~\cite{dw}, and also 
in ballistic
double magnetic Josephson junctions, provided that the thicknesses
of the magnetic layers are unequal~\cite{trifunovic}. Recently,
evidence of higher harmonics has been  observed
in Josephson junctions with spin dependent tunneling
barriers.~\cite{pal}

The interaction of  the spin current 
with the magnetization in 
layered  ferromagnetic junctions  
with multiple ferromagnets has important consequences
for memory technologies. 
Indeed, storage of  information bits depends 
on the precise relative orientation of the
magnetizations in  two $F$ layers, where
nonconserved spin currents 
reflect the mutual  torque acting on
the magnetic moments.
The corresponding spin transfer torque (STT)
can also switch
magnetizations
when a
spin-polarized electrical current 
flows perpendicular to the layers.
Spin-transfer torque is known to occur in a
very broad variety of materials, making it an attractive switching mechanism.
For equilibrium spin currents, governed by spin-polarized Andreev bound states\cite{sauls}, 
tuning the supercurrent (via $\Delta \varphi$ )
directly influences  the STT when varying the relative in-plane magnetization angle.~\cite{brou} 
The direction of supercurrent flow however is not simply related to the direction of the induced torque
that
tends to align the  magnetic moments.
The
triplet correlations
generated  
in these types of Josephson junctions (with noncollinear relative magnetizations),
can also induce spatial variations in the
spin currents responsible for the mutual torques acting on the ferromagnets.~\cite{Shomali}

When considering superconducting proximity effects, it is 
important to make sure that the self-consistency condition for the pair potential $\Delta(x)$
is fulfilled~\cite{wvh14} in order to obtain the correct physical picture.
The self-consistency condition is often neglected in the literature~\cite{liu}
mainly because it is difficult to properly  implement in the theoretical studies.
As we will show in the Sec.~\ref{methods}, a source term in the 
continuity equation for the charge currents usually arises when 
a non self-consistent superconducting order parameter is used.
More importantly, when the self-consistent condition is achieved,
the free energies of these proximity-coupled systems are properly minimized.  
This concept is crucial especially when studying Josephson junctions,
since the superconducting proximity effects are the fundamental 
mechanism behind the nontrivial charge and spin currents that
flow within these structures. Furthermore, when the solutions are
self-consistent,  the
charge conservation law is satisfied by properly accounting for the
proximity effects and transport properties for the charge
current~\cite{riedel}. When  magnetic inhomogeneity is present, the spin density
is not conserved and STT 
can arise and interact with the charge
dependent quantities.
Although  there is no continuity equation 
for the spin density, since its gradients are nonzero,
there is still a corresponding, and fundamental, 
conservation law that balances current gradients
and the STT~\cite{wvh14,linder}. 

Above, we have discussed various physical phenomenon associated with
the magnetic inhomogeneity such as triplet correlations and the generations
of STT.  
Thus understanding the interplay between spin and charge transport 
with the long-range proximity effects is an important topic
and constitutes the main goal of this work.
In this work, we therefore consider nanoscale $SFS$ Josephson junctions,
where the $F$ region contains multiple layers: 
$SFFS$ spin valves consisting of two metallic 
ferromagnetic layers separated by a non-magnetic normal metal spacer,
and trilayer $SFFFS$ junctions. In each scenario, 
a supercurrent is established via a phase difference $\Delta\varphi$
between the $S$ terminals,
and different 
 relative magnetization orientations  are considered.
The paper is organized as follows: We describe our method in Sec~\ref{methods},
and derive conservation laws for both spin and charge currents. 
Our method is based on solving the microscopic Bogoliubov-de Gennes (BdG) equations
self-consistently.
All important physical quantities, e.g., the magnetization, can be extracted or constructed 
from the self-consistent solutions. In Sec~\ref{results},
we present a detailed study of the transport properties. We conclude
with our main findings in Sec~\ref{conclusions}.

\begin{figure}
\centerline{\includegraphics[width=8.8cm]{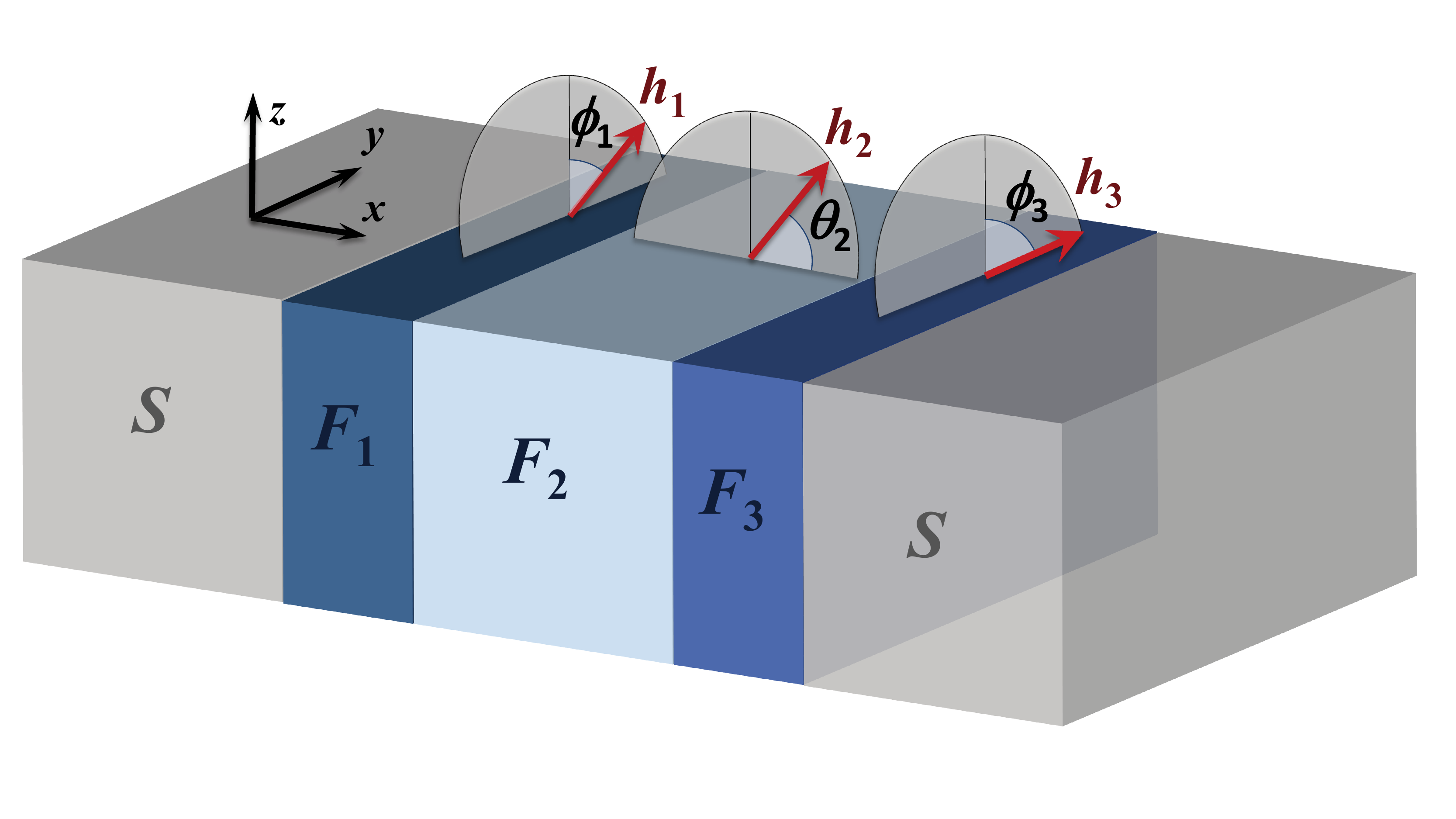}}
\caption{Schematic of the $SF_1 F_2 F_3 S$ Josephson junction considered in this paper.
A generic configuration is shown, described by the in-plane magnetization angles, $\phi_i$, 
and the out-of-plane angles, $\theta_i$ ($i=1,2,3$). The ferromagnetic exchange field ${\bm h}_i$ in each region is expressed
as ${\bm h}_i = h_i(\cos\theta_i,\sin\theta_i\sin\phi_i,\sin\theta_i\cos\phi_i)$.
 }   
\label{fig0}
\end{figure}

\section{Methods}
\label{methods}

The  general method that we use in this paper is that of 
numerical
diagonalization of the self-consistent Bogoliubov-de Gennes 
(BdG) equations. 
Since many aspects of this method have been extensively
discussed in previous work\cite{hv1,dw,wvh12,wvh14} we will only include
here a 
brief review of these points, as needed to make this paper understandable.
We will discuss in more detail additional aspects needed 
for the transport calculations described in this work.
 
The derivation of the BdG equations for
general magnetization configurations
begins with the effective BCS Hamiltonian, ${\cal H}$,
\begin{align} \label{heff}
{\cal H} =\int d^3 r\Bigl\lbrace 
\psi^\dagger ({\bm r})[  {\cal H}_e- & {\bm h} \cdot {\bm \sigma}] \psi ({\bm r})
 + 
 \Delta({\bm r}) \psi^\dagger_\uparrow({\bm r}) \psi^\dagger_\downarrow({\bm r})
\nonumber \\ &+\Delta^*({\bm r}) \psi_\downarrow({\bm r}) \psi_\uparrow({\bm r})
\Bigr\rbrace,
\end{align}
where 
$\psi({\bm r}) \equiv
(\psi_\uparrow,\psi_\downarrow)^T$ are the usual fermionic operators,  
${\cal H}_{\rm e}=-1/(2m) \bnabla^2-E_F+U({\bm r})$,
and
\mbox{\boldmath $\sigma$} denote the set of Pauli matrices.
We describe the magnetism of
the $F$ layers by  effective  Stoner exchange fields 
${\bm h}({\bm r})$ which in our case 
have components in all $(x,y,z)$ directions (see Fig.~\ref{fig0}). The
spin independent scattering potential is denoted by
$U({\bm r})$, 
and $ \Delta({\bm r})$ is the pair potential. 

To diagonalize 
the effective Hamiltonian, the field operators $\psi_\uparrow$ and
$\psi_\downarrow$ 
are expanded\cite{bdg} by means of a Bogoliubov transformation:
\begin{subequations}
\label{bv}
\begin{align}
\psi_{\uparrow}({\bm r})&=\sum_n \left(u_{n\uparrow}({\bm r})\gamma_n - v^*_{n\uparrow}({\bm r})\gamma_n^\dagger\right), \\ 
\psi_{\downarrow}({\bm r})&=\sum_n \left(u_{n\downarrow}({\bm r})\gamma_n + v^*_{n\downarrow}({\bm r})\gamma_n^\dagger\right),
\end{align}
\end{subequations}
where $u_{n\alpha}$ and $v_{n\alpha}$ are the quasiparticle 
and quasihole amplitudes, which are chosen so that the Hamiltonian
is diagonalized in terms of
the fermionic $\gamma_n$ 
operators.
Therefore, $[{\cal H},\gamma_n]=-\epsilon_n\gamma_n$ and
$[{\cal H},\gamma^\dagger_n]=\epsilon_n\gamma^\dagger_n$.
Also, the thermal expectation values involving 
$\gamma_n$ and $\gamma_n^\dagger$ are given by the usual Fermi
functions $f_n$.~\cite{song} 
The anticommutation relations for $\psi$ and $\psi^\dagger$
yield,
\begin{subequations}
\label{comm}
\begin{align}
[\psi_{\uparrow}({\bm r}),{\cal H}]&=({\cal H}_{\rm e}-h_z)\psi_{\uparrow}({\bm r})-[h_x-ih_y]\psi_{\downarrow}({\bm r})
+ \Delta({\bm r})\psi^\dagger_{\downarrow}({\bm r}), \\
[\psi_{\downarrow}({\bm r}),{\cal H}]&=({\cal H}_{\rm e}+h_z)\psi_{\downarrow}({\bm r})-[h_x+ih_y]\psi_{\uparrow}({\bm r})
-\Delta({\bm r})\psi^\dagger_{\uparrow}({\bm r}).
\end{align}
\end{subequations}

It is convenient at  this point to simplify to the quasi
one-dimensional geometry (Fig.~\ref{fig0}) of our problem.
Then, by 
inserting (\ref{bv}) into (\ref{comm}) and using the commutation
relations,
we obtain the general spin-dependent BdG equations for this
geometry,
\begin{align}
&\begin{pmatrix}  
\mathcal{H}_0 -h_z&-h_x+ih_y&0&\Delta \\
-h_x-ih_y&\mathcal{H}_0 +h_z&\Delta&0 \\
0&\Delta^*&-(\mathcal{H}_0 -h_z)&-h_x-ih_y \\
\Delta^*&0&-h_x+ih_y&-(\mathcal{H}_0+h_z) \\
\end{pmatrix}
\begin{pmatrix}
u_{n\uparrow}(x)\\u_{n\downarrow}(x)\\v_{n\uparrow}(x)\\v_{n\downarrow}(x)
\end{pmatrix}
\nonumber \\ &=\epsilon_n
\begin{pmatrix}
u_{n\uparrow}(x)\\u_{n\downarrow}(x)\\v_{n\uparrow}(x)\\v_{n\downarrow}(x)
\end{pmatrix},
\label{bogo}
\end{align}
where  $\epsilon_n$  are the quasiparticle energies, 
$x$ is normal to the layers, which lie in the $y-z$ plane 
(see Fig.~\ref{fig0}), 
and $\Delta(x)$ is the pair potential, to be found self consistently
as explained below.
Here 
the single particle Hamiltonian ${\cal H}_0$ is written, 
\begin{equation}
{\mathcal H}_0 = 
\frac{1}{2m}\left(-\frac{\partial^2}{\partial x^2}+k_y^2+k_z^2 \right)-E_F +
U(x). 
\end{equation}
The components of the exchange field ${\bm h}$ in each of the $F$ layers
take the form:
\begin{align}
\label{fields}
{\bm h}_i=h(\cos\theta_i,\sin\theta_i\sin\phi_i,\sin\theta_i\cos\phi_i), 
\end{align}
where $i$ denotes one of the magnetic
layers. We will assume that the magnitude of the exchange field
is 
the same in all magnetic layers, and that it vanishes elsewhere.
The angles $\theta_i$ and $\phi_i$ will in general be taken to
vary from layer to layer.

One obtains in the usual way\cite{bdg}
the self-consistency condition for the pair potential, using
$\Delta({\bm r}) = g\langle \psi_\downarrow({\bm r})\psi_\uparrow 
({\bm r})\rangle$. Here $g$ is the superconducting coupling constant, which
vanishes in the non-$S$ layers. In our geometry we find, after using
the Bogoliubov transformation and 
making use of
the appropriate averages such
as $\langle \gamma^\dagger_n \gamma_n \rangle= f_n$,
the
pair potential can be expressed
 in terms of the quasiparticle amplitudes 
as an appropriate sum over states:
\begin{equation}  
\label{del2} 
\Delta(x) = \frac{g}{2}{\sum_{n}}^\prime
\left[u_{n\uparrow}(x)v^*_{n\downarrow} (x)+
u_{n\downarrow}(x)v^*_{n\uparrow} (x)\right]\tanh(\epsilon_n/2T), 
\end{equation} 
where the prime on the sum indicates that only those states 
that have energies within 
a ``Debye energy", $\omega_D$, are included.

The problem is then solved  iteratively: the potential  
is initially taken to be  $\Delta_0$ in the first 
$S$ layer and $\Delta_0 \exp(i\Delta\varphi)$ 
in the second $S$ layer, where $\Delta_0$ is the initial guess
of the magnitude for the pair amplitudes. 
The Hamiltonian is then numerically
diagonalized and the new pair potential is found 
via Eq.~(\ref{del2}). Iteration
is continued until convergence. 
The details of the procedure, including the way
to ensure, in the Josephson calculations, that the phase difference between
the right and left ends of the sample remains $\Delta\varphi$, is explained
in Appendix \ref{appB}.


Equilibrium and transport properties in the structures 
considered are strongly influenced by the existence of ``odd'' triplet
pairs. The existence of such pairs is allowed by
conservation laws since, unless all of the $F$ layers have
magnetizations along the same direction, the total spin of the Cooper
pairs is not a conserved quantity. Because of the Pauli principle,
these $s$-wave triplet pairs must have wavefunctions odd in frequency\cite{brz}
or equivalently\cite{hbvprl,hvb08} in time. 
Within the BdG framework, the 
time formulation is much more convenient. Accordingly,
we will describe the triplet pair correlations via 
the following 
amplitude functions, in terms of the field operators:
\begin{subequations}
\label{pa}
\begin{align}
{f_0}({\bm r},t) =& \frac{1}{2}[\langle \psi_{\uparrow}({\bm r},t) \psi_{\downarrow} 
({\bm r},0)\rangle+
\langle \psi_{\downarrow}({\bm r},t) \psi_{\uparrow} ({\bm r},0)\rangle],\\
{f_1}({\bm r},t) =& \frac{1}{2}[\langle \psi_{\uparrow}({\bm r},t) \psi_{\uparrow} 
({\bm r},0)\rangle -\langle \psi_{\downarrow}({\bm r},t) \psi_{\downarrow} 
({\bm r},0)\rangle].
\end{align}
\end{subequations}
Taking  the quantization axis  along the $z$ direction, the
triplet amplitudes, $f_{0}$ and $f_{1}$, can be rewritten\cite{hbvprl,hvb08}
in terms of the
quasiparticle amplitudes:
\begin{align}
\label{f0}
f_{0} &=  1/2\sum_{n}(g_n^{\uparrow\downarrow}-g_n^{\downarrow\uparrow}) \zeta_n(t), \\
\label{f1}
f_{1} &  =1/2\sum_{n} (g_n^{\uparrow\uparrow}+g_n^{\downarrow\downarrow})\zeta_n(t),
\end{align}
where $\zeta_n(t) \equiv \cos(\epsilon_n t)-i\sin(\epsilon_n
t)\tanh(\epsilon_n/2 T)$, and we define 
$g_{n}^{\sigma\sigma'} \equiv u_{n \sigma} v^{\ast}_{n \sigma'}$.
It is sometimes necessary to evaluate the triplet amplitudes along
a different spin axis. For example, one may wish to
use the direction of the local magnetization (defined below)
as the axis of quantization. To do so one  rotates the
quantization axis so that it is aligned with the local magnetization
direction using the spin rotation matrices discussed
in Appendix \ref{appA}).

We will consider here spin currents, as well as charge
currents. In our structures spin transport is influenced by the 
leakage of magnetism out of the $F$ layers and into
the superconductors. This can be characterized
by the local magnetization ${\bm m}({\bm r})$, 
\begin{align} \label{mag}
{\bm m}({\bm r})  =-\mu_B\, \langle {\bm \eta}({\bm r})   \rangle,
\end{align}
where ${\bm \eta}({\bm r}) $ is the spin density operator,
\begin{align} \label{spinop}
{\bm \eta}({\bm r})  = \psi^\dagger({\bm r})  {\bsigma} \psi({\bm r}) ,
\end{align}
and $\mu_B$  the Bohr magneton. 
For our quasi-1D geometry,  
we can rewrite the components of ${\bm m}$
in terms of the quasiparticle amplitudes: 
\begin{align}
m_x(x)&=- \mu_B\sum_n \Bigl\lbrace \Bigl[u^*_{n\uparrow}(x)u_{n\downarrow}(x)+u^*_{n\downarrow}(x)u_{n\uparrow}(x)  \Bigr]f_n \nonumber \\
&-\Bigl[v_{n\uparrow}(x)v^*_{n\downarrow}(x)+v_{n\downarrow}(x)v^*_{n\uparrow}(x) \Bigr](1-f_n) \Bigr\rbrace. \label{mx} \\
m_y(x)&=- i\mu_B\sum_n \Bigl\lbrace \Bigl[u_{n\uparrow}(x)u^*_{n\downarrow}(x)-u_{n\downarrow}(x)u^*_{n\uparrow}(x)  \Bigr]f_n \nonumber \\
&+\Bigl[v_{n\uparrow}(x)v^*_{n\downarrow}(x)-v_{n\downarrow}(x)v^*_{n\uparrow}(x) \Bigr](1-f_n) \Bigr\rbrace. \label{my} \\
m_z(x)&=-\mu_B\sum_n \Bigl\lbrace \Bigl[|u_{n\uparrow}(x)|^2 - |u_{n\downarrow}(x)|^2 \Bigr]f_n \nonumber \\
&+  \Bigl[|v_{n\uparrow}(x)|^2 - |v_{n\downarrow}(x)|^2 \Bigr](1-f_n) \Bigr \rbrace. \label{mz}
\end{align}


We now turn to the appropriate expressions for the currents.
As stressed in the Sec.~\ref{introduction}, one
needs to carefully establish proper conservation laws 
when discussing the transport properties of the system~\cite{kadanoff}. 
We first discuss the charge supercurrent. 
In our geometry the charge current has only one component, $J_x$
which depends  on the $x$ coordinate.
In the absence of an external magnetic field,
the total charge 
current, $J_x(x)\equiv J_{x\uparrow}(x)+J_{x\downarrow}(x)$,  
is found from the standard quantum mechanical expression, 
${J}_x = (e/m) \langle \psi^\dagger p_x \psi \rangle$.
This leads to
the result:
\begin{align} \label{jay}
{J}_{x\sigma}(x) = \frac{e}{2m} \left\langle  -i \psi_\sigma^\dagger \frac{\partial}{\partial x}
\psi_\sigma+ i \left(\frac{\partial}{\partial x} \psi_\sigma^\dagger\right) \psi_\sigma \right\rangle.
\end{align}
This expression for the current  ensures,
together with the self consistency  condition, that
charge conservation is satisfied, that  is, $d J_x/ dx=0$ 
in the steady state.~\cite{wvh14,dw,sols} 
It is of course convenient 
numerically to rewrite the expression for the
current in terms of the  calculated 
quasiparticle amplitudes and energies.
After inserting the Bogoliubov transformations in Eq.~(\ref{bv}),
we can write the total charge current, as given by Eq.~(\ref{jay}) 
summed over spins
as:
\begin{align}
J_{x}(x)&=\frac{2e}{m}\sum_{n,\sigma} {\rm Im }\Bigl[u_{n \sigma} \frac{\partial u^{*}_{n \sigma}}{\partial x} f_n+
v_{n \sigma}\frac{\partial v^{*}_{n \sigma}}{\partial x} \left(1-f_n\right) \Bigr].
\label{cur}
\end{align}
One can verify once again
the conservation law by
taking the divergence of the current in Eq.~(\ref{cur}) and using the 
BdG equations (\ref{bogo}), to find:~\cite{dw,wvh14}
\begin{align}
\frac{\partial {J_x(x)}}{\partial x}
=2 e {\rm Im} \left \lbrace \Delta(x) \sum_n 
[u^*_{n \uparrow} v_{n\downarrow} +u^*_{n\downarrow} v_{n \uparrow}] 
\tanh\left(\frac{\epsilon_n}{2T} \right) \right \rbrace.
\end{align}
When the self-consistency condition is satisfied, the right hand side 
vanishes, and 
charge is properly conserved. If the self-consistency condition is not 
strictly satisfied, 
the terms on the right side act effectively as sources or sinks of 
current.\cite{sols,wvh14,dw}
We  will consider large $S$ contacts with the amplitude
and phase of the order parameter determined self-consistently
except near
the sample edges (see Appendix \ref{appB}) where sources and sinks of
charge exist via the implicit
external electrodes. This gives the necessary charge 
conservation condition in the
region of interest.
We emphasize here that
with self-consistent solutions,
we are able to correctly determine the effect of triplet
correlations on both the charge and spin transport. 

The extension of the above considerations to spin
transport is
relatively straightforward.\cite{brou,wvh14} 
As in the case of the charge density, the Heisenberg picture is  
utilized  to determine the time evolution of the spin density, 
${\bm \eta}({\bm r},t)$,
\begin{align} \label{scom}
\frac{\partial}{\partial t} \langle {\bm \eta}({\bm r},t) \rangle = i \langle 
[{\cal H},{\bm \eta}({\bm r},t)] \rangle,
\end{align}
where ${\bm \eta}$ is given in Eq.~(\ref{spinop}).
The associated continuity equation now reads,
\begin{align} \label{scon}
\frac{\partial}{\partial t} \langle {\bm \eta}({\bm r},t) \rangle + \frac{\partial {\bm S}}{\partial x} &= 
{\bm \tau}+{\cal J}_S,
\end{align}
where ${\bm S}$ is the spin current which  in our 
geometry is a vector (in general it is a tensor).
The spin-transfer torque, ${\bm \tau}$, is given by: 
\begin{align} \label{stt}
{\bm \tau}=-i \langle \psi^\dagger({\bm r}) [{\bm h}\cdot{\bm \sigma},{\bm \sigma}] \psi ({\bm r})\rangle 
=2 \langle \psi^\dagger({\bm r}) [{\bm \sigma}\times{\bm h}] \psi ({\bm r})\rangle.
\end{align}
The  ${\cal J}_S$ term has components,
\begin{align}
{\cal J}_{Sx} &= 2{\rm Im} \left\lbrace \Delta 
\Bigl\langle \psi^\dagger_{\downarrow}({\bm r}) \psi^\dagger_{\downarrow}({\bm r})  
-\psi^\dagger_\uparrow ({\bm r}) \psi^\dagger_\uparrow ({\bm r}) \Bigr\rangle\right\rbrace, \\
{\cal J}_{Sy} &= 2{\rm Re} \left\lbrace \Delta 
\Bigl\langle \psi^\dagger_{\downarrow}({\bm r}) \psi^\dagger_{\downarrow}({\bm r})  
+\psi^\dagger_\uparrow ({\bm r}) \psi^\dagger_\uparrow ({\bm r}) \Bigr\rangle\right\rbrace.
\label{source2}
\end{align}
For the $s$-wave superconductors considered in this paper, 
we have ${\cal J}_S=0$ by virtue of the Pauli principle, since only
equal time  correlations are
involved. 
Thus in the absence  spin transfer torque, 
we have 
${\partial {\bm \eta}}/{\partial t}+{\partial {\bm S}}/{\partial x}= 0$.
However, in general\cite{wvh14}
spin transfer torque is present and, in the steady state, 
the derivatives of ${\bm S}$ with respect to $x$ do not vanish.

The expression for the 
spin-current, ${\bm S}$, is found from taking the commutator 
in Eq.~(\ref{scom}) and using Eq.~(\ref{heff}): 
\begin{align}
{\bm S} = -\frac{i}{2m} \Bigl \langle \psi^\dagger({\bm r}) {\bm \sigma} \frac{\partial \psi({\bm r})}{\partial x}  
- \frac{\partial  \psi^\dagger({\bm r})}{\partial x} {\bm \sigma} \psi({\bm r})
\Bigr \rangle,
\end{align}
where we recall that the vector 
${\bm S}$ 
represent spin current flowing along the $x$ direction for
our quasi-one-dimensional systems. 
We can now expand each spin component of the spin current in terms of 
the quasiparticle amplitudes to obtain: 
\begin{align}
&{\bm S}_x = -\frac{i}{2m}\sum_n \Biggl\lbrace f_n\Bigl[u_{n\uparrow}^* \frac{\partial u_{n \downarrow}}{\partial x}+
u_{n\downarrow}^* \frac{\partial u_{n \uparrow}}{\partial x}-
u_{n\downarrow} \frac{\partial u^*_{n \uparrow}}{\partial x}-
u_{n\uparrow}\frac{\partial u^*_{n \downarrow}}{\partial x} \Bigr ] \nonumber \\
&-(1-f_n)
\Bigl[v_{n\uparrow} \frac{\partial v^*_{n \downarrow}}{\partial x}+
v_{n\downarrow} \frac{\partial v^*_{n \uparrow}}{\partial x}-
v^*_{n\uparrow} \frac{\partial v_{n \downarrow}}{\partial x}-
v^*_{n\downarrow} \frac{\partial v_{n \uparrow}}{\partial x} \Bigr ] \Biggr\rbrace, \\
&{\bm S}_y = 
-\frac{1}{2m}\sum_n \Biggl\lbrace f_n\Bigl[u_{n\uparrow}^* \frac{\partial u_{n \downarrow}}{\partial x}-
u_{n\downarrow}^* \frac{\partial u_{n \uparrow}}{\partial x}-
u_{n\downarrow} \frac{\partial u^*_{n \uparrow}}{\partial x}+
u_{n\uparrow}\frac{\partial u^*_{n \downarrow}}{\partial x} \Bigr ] \nonumber \\
&-(1-f_n)
\Bigl[v_{n\uparrow} \frac{\partial v^*_{n \downarrow}}{\partial x}-
v_{n\downarrow} \frac{\partial v^*_{n \uparrow}}{\partial x}+
v^*_{n\uparrow} \frac{\partial v_{n \downarrow}}{\partial x}-
v^*_{n\downarrow} \frac{\partial v_{n \uparrow}}{\partial x} \Bigr ]\Biggr\rbrace, \\
&{\bm S}_z =
-\frac{i}{2m}\sum_n \Biggl\lbrace f_n\Bigl[u_{n\uparrow}^* \frac{\partial u_{n \uparrow}}{\partial x}-
u_{n\uparrow} \frac{\partial u^*_{n \uparrow}}{\partial x}-
u^*_{n\downarrow} \frac{\partial u_{n \downarrow}}{\partial x}+
u_{n\downarrow}\frac{\partial u^*_{n \downarrow}}{\partial x} \Bigr ] \nonumber \\
&-(1-f_n)
\Bigl[-v_{n\uparrow} \frac{\partial v^*_{n \uparrow}}{\partial x}+
v^*_{n\uparrow} \frac{\partial v_{n \uparrow}}{\partial x}+
v_{n\downarrow} \frac{\partial v^*_{n \downarrow}}{\partial x}-
v^*_{n\downarrow} \frac{\partial v_{n \downarrow}}{\partial x} \Bigr ]\Biggr\rbrace.
\end{align}
In the case of $F$ layers with uniform magnetization,
there is no net spin current. The introduction of an
inhomogeneous magnetization texture however
results in a net spin current imbalance 
that is finite\cite{wvh14}  even in 
the absence of a Josephson current. 
This will be discussed in greater detail
below.

To compute the spin transfer torque, it is useful to
express it in terms of the quasiparticle amplitudes.
A convenient approach involves
directly taking the expectation values of Eq.~(\ref{stt}):
\begin{equation} \label{tau1}
{\bm \tau}
=2\langle \psi^\dagger({\bm r}) {\bm \sigma}\psi ({\bm r})\rangle\times{\bm h}
= -\frac{2}{\mu_B} {\bm m} \times {\bm h},
\end{equation}
where we have used  Eq.~(\ref{mag}). 
The magnetization components are given in Eqs.~(\ref{mx})-(\ref{mz}).
Since the exchange field ${\bm h}$ is  prescribed, it 
is the self consistently calculated magnetization that determines the
torque acting on the ferromagnet layers.
Equivalently,
one can use
the continuity equation in the steady state
to determine the torque transfer by
evaluating the
derivative of the spin current as a function of position:
\begin{align} \label{sx1}
{\bm \tau}_{i}=
\frac{\partial {\bm S}_i}{\partial x}.
\end{align}
It is however safer to evaluate both sides of Eq.~(\ref{sx1}) 
independently and use this equation as a consistency check.
We have performed extensive numerical checks 
 using this procedure. In most of the results presented
we have
calculated the torques using  Eq.~(\ref{tau1}), 
thus avoiding the numerical derivatives that 
arise when using the right side of Eq.~(\ref{sx1}).
Additional physical insight can be gained by  integrating
Eq.~(\ref{sx1})   over a particular region, e.g.,
$F_1$:
\begin{align} \label{sx2}
 {\bm S}_{x}(b)- {\bm S}_{x}(a)=
\int_{F_1} dx  { \tau}_x= \tau_{x,{\rm tot}}    ,
\end{align}
which means that the change in spin current
through $F_1$   (from $x=a$ to $x=b$)
is equivalent to
the net torque acting within those boundaries.

\section{Results}
\label{results}

The results of our systematic investigations
are presented   below in terms of convenient 
dimensionless quantities. 
Our choices are as follows:
all length scales, including the 
position $X\equiv k_F x$, and widths 
$D_{Fi}\equiv k_F d_{Fi}$ ($i=1,2,3$) 
are normalized 
by the Fermi wavevector, $k_F$.
For the
superconducting
correlation length $\xi$ we choose the 
value $k_F \xi = 100$, and
the computational region occupied by the $S$ electrodes 
corresponds to a width of $8\xi$  (see Appendix \ref{appA}
for numerical details).
All temperatures are measured in units of $T_{c0}$, 
the transition temperature of bulk
$S$ material, and
we consider  the low temperature regime, $T/T_{c0}=0.01$. 
Energy scales are normalized by the Fermi energy, $\varepsilon_F$,
including the Stoner field interaction ${\bm h}$ and
the energy cutoff, $\omega_D$, used in the 
self-consistency condition, Eq.~(\ref{del2}). 
The latter is set at $0.04$: results
are independent of this cutoff choice.
As mentioned above, the strength of the magnetic exchange 
fields, $h$ is taken to be the same in both
magnets: we set its dimensionless value 
to a representative 
$h=0.1$.
We vary the orientation angles of the magnetic
exchange field in each of the $F$ regions,
depending on the quantity being studied.
The magnetization is normalized by
 $\mu_B {n}_e$, 
  where $n_e$ is the
 electron density, $n_e=k_F^3/(3\pi^2)$.
The normalization $\tau_0$ for the torque follows
from the normalizations for $h$ and ${\bm m}$ and Eq.~(\ref{tau1}). 
When presenting results for the currents, 
we normalize the charge current densities $J_x$ 
by $J_0$, where
$J_0 \equiv e n_e v_F$, and $v_F = k_F/m$ is the Fermi velocity.
All three components of the 
spin current ${\bm S}$ are normalized similarly, by 
the quantity $S_0$, where $S_0$ involves the
normalization of ${\bm m}$ and a factor of  $n_e v_F$. 
The interface scattering $U(x)$  is represented by delta functions
of strength $H$  
at all the interfaces. The 
corresponding dimensionless
parameter is $H_B\equiv H/v_F$.
The self-consistency of the pair potential that characterizes an accurate representation
of the Cooper pair correlations throughout the system, 
is associated with the proximity effects and depends to varying degrees on 
the parameters outlined above. In some cases the
dependence is rather obvious: for instance, large $H_B$ results 
in weaker proximity effects. In other cases it is more intricate and
will be analyzed more carefully.

\begin{figure}
\centerline{\includegraphics[width=8.8cm]{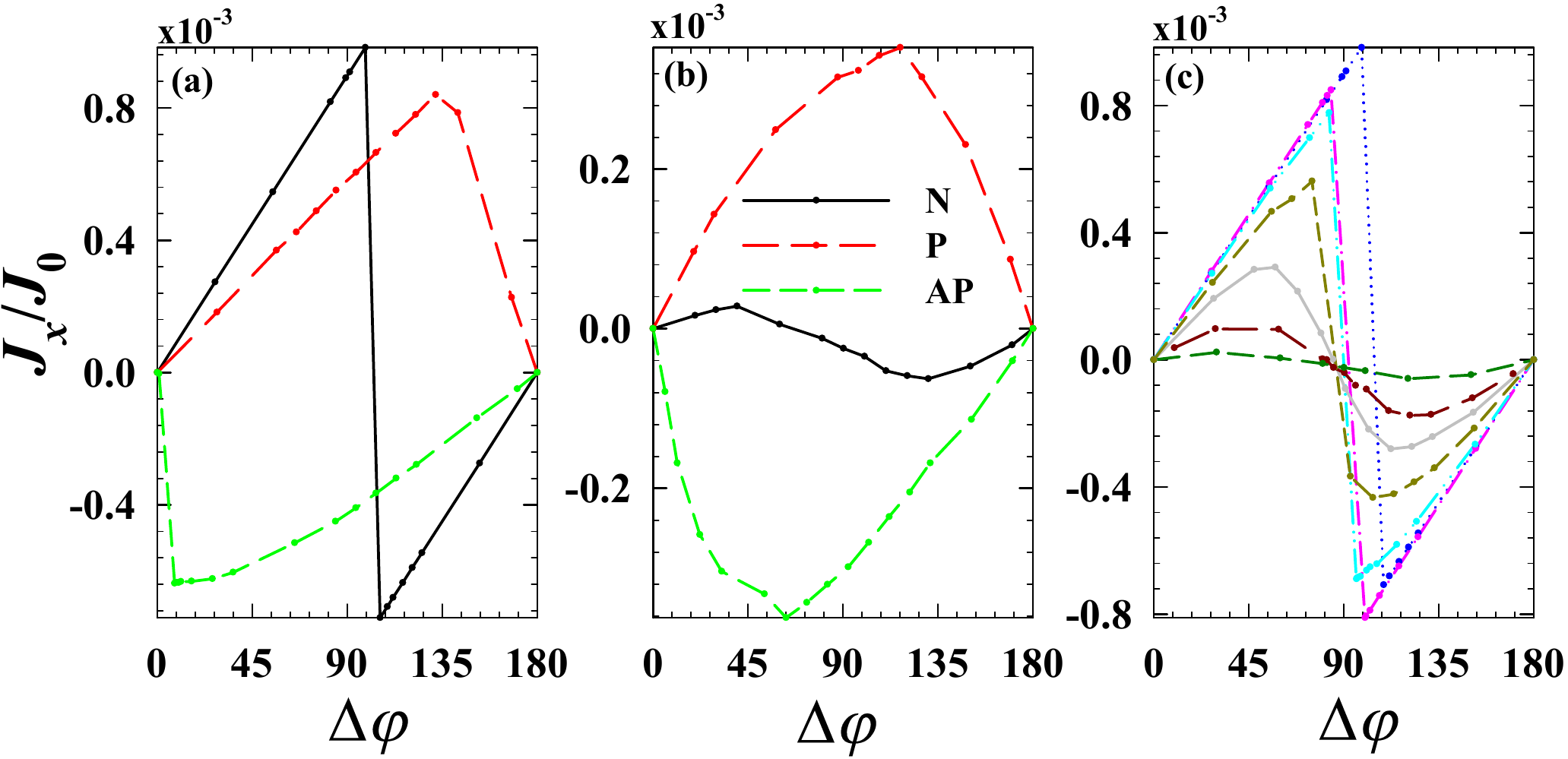}}
\caption{Normalized 
(see text) Josephson current 
versus phase difference, 
$\Delta\varphi$, 
for a $S F_1 F_2 S$ structure with $D_{F1}=10$, 
$D_{F2}=100$, and $h=0.1$. 
For panels (a) and (b), the legend
in (b) labels the relative in-plane magnetization orientations:
parallel (P), antiparallel (AP), or  normal (N).
Two interface scattering strengths are considered: (a)  $H_B=0$, and  
(b) $H_B=1$.
In panel (c), the  magnetization orientations are fixed in the  normal 
configuration ($\phi_1=0$, $\phi_2=90^{\circ}$), 
and the interface scattering is varied as $H_B=0,0.2,0.5,0.6,0.7,0.8,1$ (in descending order of peaks).
 }   
\label{fig1}
\end{figure}

\subsection{Current-Phase relations}
\label{cpr}
We begin by showing
our results for the self-consistent current phase relations in a simple 
double layer ferromagnet Josephson junction.
At the interfaces between the $F$ and $S$ regions,
quasiparticles undergo Andreev and
conventional reflections.\cite{radovic2,radovic1,beenaker2,beenaker1}
The superposition of
these waves in the $F$ regions results in subgap bound states that contribute,
together with the continuum states,
to the total current flow.
In Fig.~\ref{fig1}, we show the supercurrent as a function of phase 
difference (current phase relation CPR) for 
two ferromagnets of unequal width:
$D_{F1}=10$ and $D_{F2}=100$.
This asymmetric choice of  widths helps ensure\cite{trifunovic,ilya}
that equal-spin triplet correlations are generated
in the  system.
The angular parameters in this figure are fixed at $\theta_1=90^\circ$,
and $\theta_2=90^\circ$,
corresponding to in-plane magnetization orientations.
The $F_1$ layer has its magnetization aligned in the $z$ direction 
($\phi_1=0^\circ$).
The first two panels, (a) and (b), display
three different relative in-plane magnetization configurations
in the $F_2$ layers: parallel (P)  ($\phi_2=0^\circ$), antiparallel (AP) 
($\phi_2=180^\circ$), and normal (N)\cite{convention} ($\phi_2=90^\circ$).
Two different strengths of the interface scattering parameter are considered. 
In (a) there is no interface scattering ($H_B=0$), 
while in (b), a rather high rate of
scattering is present, with $H_B=1$.
The CPR for the  collinear configurations (P or AP) 
possesses the conventional $2\pi$ periodicity,
and the supercurrent flows oppositely for the two alignments. 
When the relative magnetizations are orthogonal to each another,
the CPR becomes $\pi$ periodic as revealed by the sawtooth-like pattern in (a)
or the more sinusoidal behavior in (b), both of which change sign
at $\Delta\varphi\approx 90^\circ$.
This is a consequence of the emergence of 
equal-spin triplet correlations\cite{2nd_hrmnc_1,
2nd_hrmnc_2,2nd_hrmnc_3,alidoust_spn,buzdin1,zareyan}
that are absent when the exchange fields in the ferromagnets 
point along the same direction.
When strong interface scattering is present, Fig.~\ref{fig1}(b) 
shows that the $\pi$ periodic CPR (N case) is substantially diminished,
relative to the P or AP cases. This is because the proximity
effect is weakened, with a resulting reduction of the 
associated equal-spin triplet correlations. 
To further examine the effects that interface scattering has on this 
$\pi$-periodic supercurrent,
we consider in Fig.~\ref{fig1}(c), the
same  $S F_1 F_2 S$ junction with varying degrees of scattering 
strengths $H_B$
and with the relative in-plane magnetizations 
fixed and
orthogonal to each another ($\phi_2=90^\circ$).
Increasing $H_B$ clearly leads
to a crossover in the CPR from a sawtooth to sinusoidal
form and to a marked reduction
of the supercurrent flow. As this occurs, 
the phase difference $\Delta \varphi$ yielding the critical current density
also declines.

\begin{figure}
\centerline{\includegraphics[width=8.8cm]{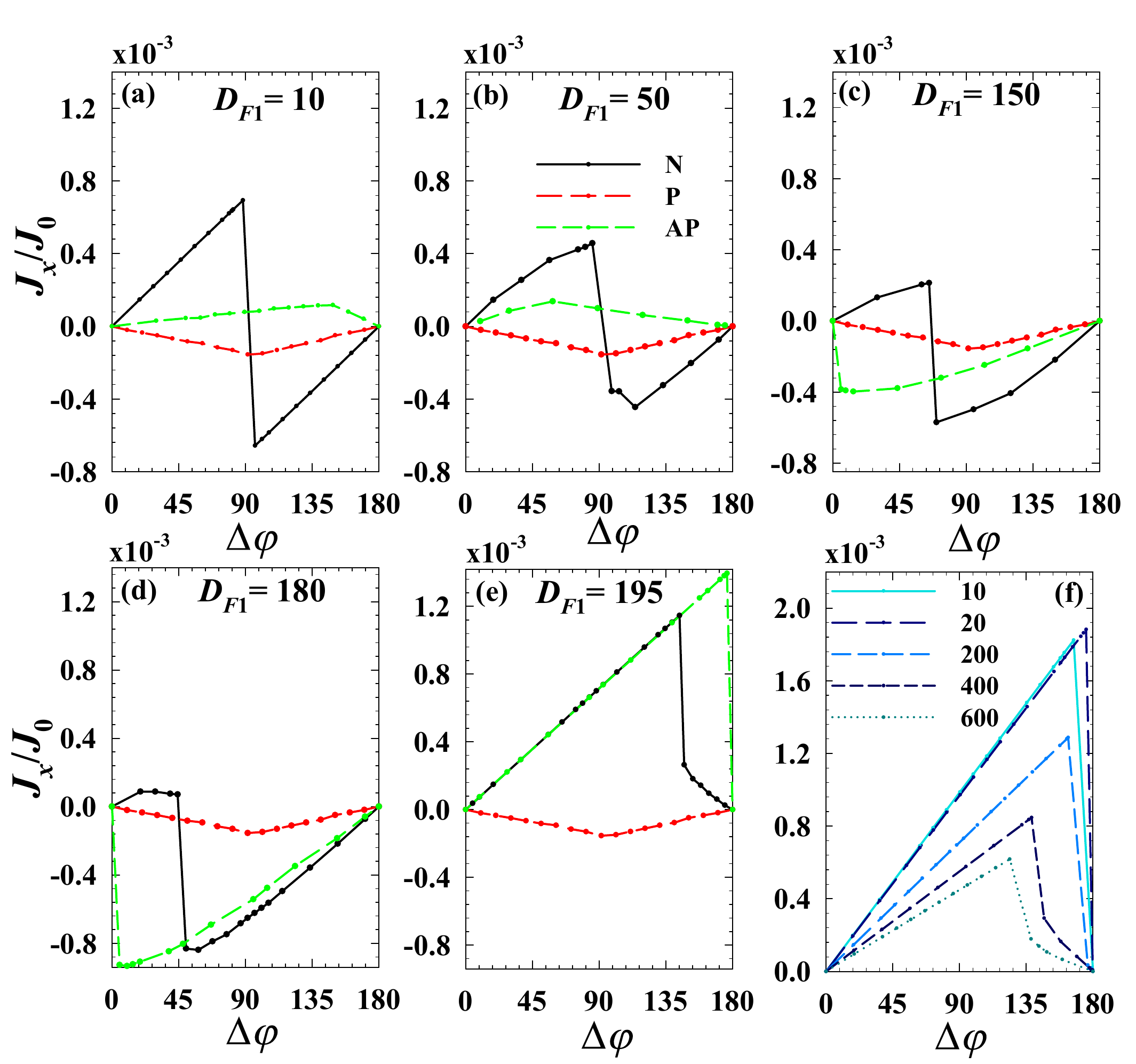}}
\caption{Normalized Josephson current 
versus phase difference, $\Delta\varphi$ 
for a $SF_1F_2S$ structure with no 
interface scattering, $H_B=0$.
In panels (a)-(e) various
widths of the $F$ regions are considered.
Each panel is labeled by  the width $D_{F1}$ of the first
ferromagnet layer $F_1$,
with
the constraint that the sum  $D_{F1}+D_{F2}=390$.
The legend in (b) labels the relative magnetization orientation: 
parallel (P),  antiparallel (AP),  and normal (N).
In (f) the magnetization orientation 
is in the N configuration, $D_{F1}$ and $D_{F2}$ are equal, and their individual
widths are given in the legend.  
 }   

\label{fig2}
\end{figure}
We next consider, in Fig.~\ref{fig2}, the effect on the CPR
when changing $D_{F1}$ and $D_{F2}$ in a $S F_1 F_2 S$ 
structure. 
Panels (a)-(e) label the various widths
$D_{F1}$ considered.
We keep the total junction length
in which supercurrent flows constant, 
i.e.,
$D_F\equiv D_{F1}+D_{F2}$, is fixed to
a representative value
of $D_F=390$. 
In these panels, we consider 
the same three orthogonal magnetization orientations 
as in Fig.~\ref{fig1}.
One sees 
that, in the N configuration, 
the structure with the greatest geometric asymmetry, 
as given by the ratio $D_{F1}/D_{F2}$, tends to have
a more pronounced superharmonic CPR relative to the P and AP collinear
configurations.
Upon increasing the width $D_{F1}$, 
the jagged sawtooth peaks at $D_{F1}=10$ become smoothed
and the $\pi$-periodic CPR 
is reduced substantially.
Eventually in panel (e),
where both $F$ widths are equal, 
the current no longer undergoes a sign change and
the additional harmonics in the CPR 
that previously reflected the existence of equal-spin triplet correlations
are now drastically modified.
Remarkably, for this situation,
over
most of the $\Delta\varphi$  range, the
linear variations of the currents in the orthogonal and antiparallel configurations 
tend to overlap.
Moreover, it is also evident that
the currents  in the N and AP configurations that flow opposite to the currents 
in the 
magnetically uniform P case, are substantially greater.
The observed reversal of current direction for a given magnetic configuration 
when changing the $F$ widths 
 is a direct consequence of the damped oscillatory
behavior of the singlet and triplet correlations, as will be discussed below (see also Fig.~\ref{fig8}).
By comparing Fig.~\ref{fig2}(a) with Fig.~\ref{fig1}(a) and noting that
the only difference between them is the thickness of the second magnet,
one finds that the magnitude of the supercurrent for the N case 
with thicker $F_2$ does not drop significantly,  as it does in the 
P and AP configurations.
Such behavior is 
a signature of the long-range nature of the equal-spin triplet 
correlations.
To show further the effects of increased width on the supercurrent,
we consider in Fig.~\ref{fig2}(f)
the
symmetric geometry configuration for several equal
 $F$ layer widths $D_F$. 
As shown in the legend,
a broad range of $D_F$ are considered.
It is clear from the figure that
the equal-spin triplet
correlations are strongly impacted,  
and
only  the  $2\pi$-periodic supercurrent arises. 
By increasing the ferromagnet widths,
the rate at which the current changes with $\Delta\varphi$
tends to decline for wider junctions,
and the ``critical" phase difference where the current  is suddenly reduced
becomes smaller. 
In this panel the ferromagnetic region 
is no longer constrained to have the same total width, hence 
increasing $D_F$ reduces the overall current flow. 

\begin{figure}
\centerline{\includegraphics[width=8.8cm]{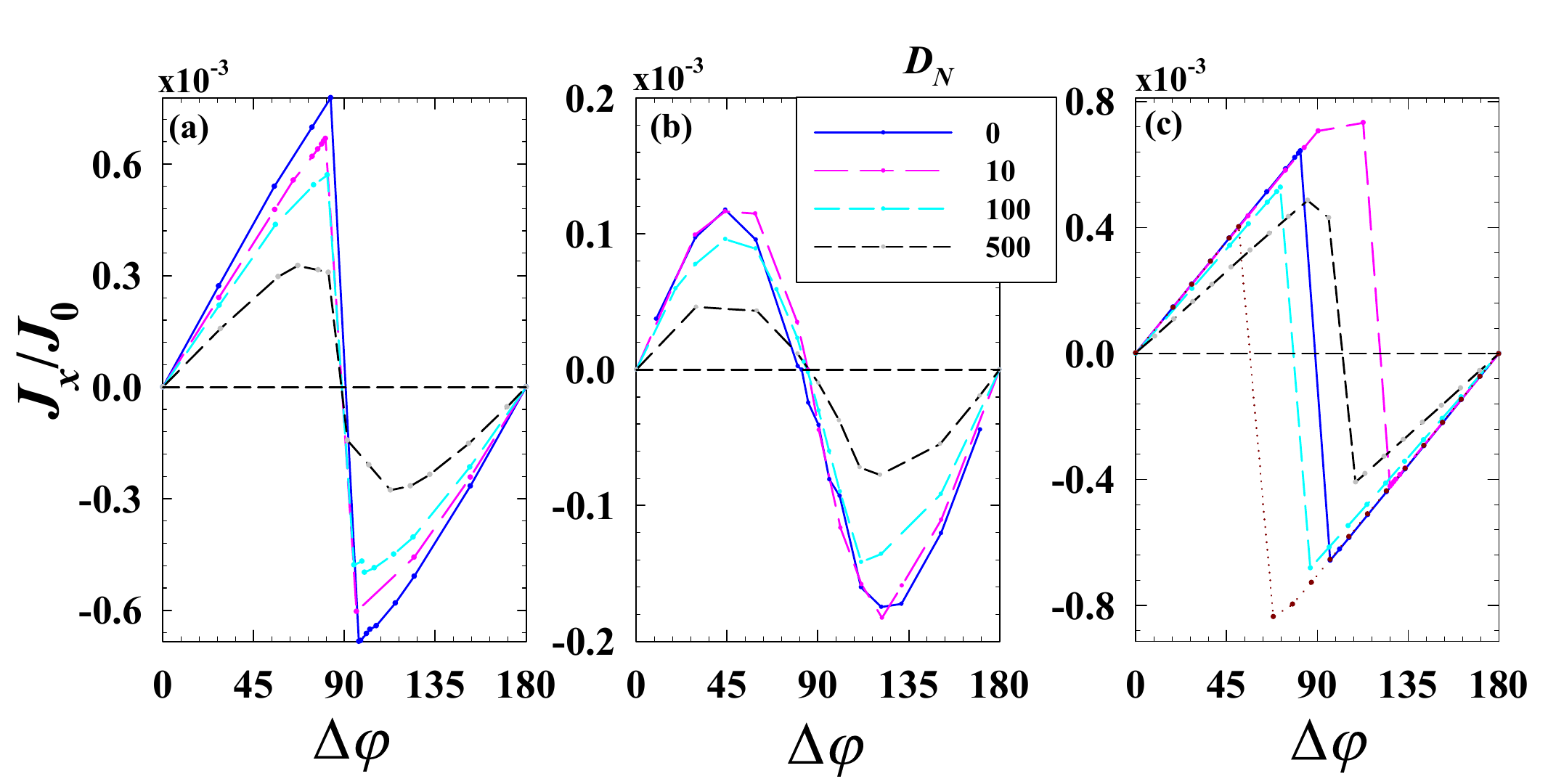}}
\caption{Normalized Josephson current 
versus  $\Delta\varphi$ 
for a $SF_1 NF_2S$ structure. The legend labels
the $N$ spacer widths, $D_N$. 
The relative in-plane magnetization angle
between the two $F$ layers is $90^\circ$.
For panels (a) and (b),
the ferromagnetic layers $F_1$ and $F_2$
have widths $D_{F1}=10$, and $D_{F2}=100$. 
In panel (a) $H_B=0.5$, in panel  (b) $H_B=0.8$, and 
in panel (c), $H_B=0$, with 
ferromagnet  widths $D_{F1}=10$, and $D_{F2}=380$. 
In panel (c) 
the
additional dotted curve
corresponds to $D_N=5$, illustrating the sensitivity of 
the current phase relation
to $D_N$.
 }   
\label{fig4}
\end{figure}
After these examples of two-magnet Josephson junctions,
we now consider a trilayer junction, where a nonmagnetic normal metal 
``spacer"
separates the two ferromagnets. Such spacers are 
often needed experimentally when it is wished\cite{ilya} to
rotate the magnetization in one magnet only.
To focus on the  case where the CPRs have
important  additional harmonics, we keep
the in-plane mutual magnetizations 
orthogonal (with $\phi_2=90^\circ$, and $\theta_2=90^\circ$).
In Figs.~\ref{fig4}(a) and (b), 
the ferromagnet widths are set
at $D_{F1}=10$ and $D_{F2}=100$, while the width of the normal metal spacer, 
$D_N$, varies 
from zero to 500 (corresponding to $0\le d_N/\xi\le 5$)
as indicated in the legend.
The interface scattering parameter has the value $H_B=0.5$ for the curves 
shown in (a),
whereas in (b) we have $H_B=0.8$.
In either case, the effect of increasing the spacer width is to
cause an overall reduction in the supercurrent flow. For 
large interface scattering, 
the CPR  becomes less sensitive to
$D_N$, as seen by comparing (a) and (b).
A wider junction with $H_B=0$ is shown in (c), with ferromagnet widths 
$D_{F1}=10$ and $D_{F2}=380$. 
The details of the 
usual sawtooth 
CPR reveals that in these cases, the supercurrent flow can be quite 
sensitive to the spacer width,
as it abruptly
reverses direction at considerably different phase differences, 
for incremental changes in $D_N$. In this panel, 
results for an additional small
value of spacer thickness, $D_N=5$ are also shown.
Since the $\pi$-periodic CPR is closely related to 
the generation of the equal-spin triplet correlation,
one can infer from Fig.~\ref{fig4} that the introduction
of an additional non-magnet metallic layer 
can quantitatively changes the transport properties.


\begin{figure}
\centerline{\includegraphics[width=8.8cm]{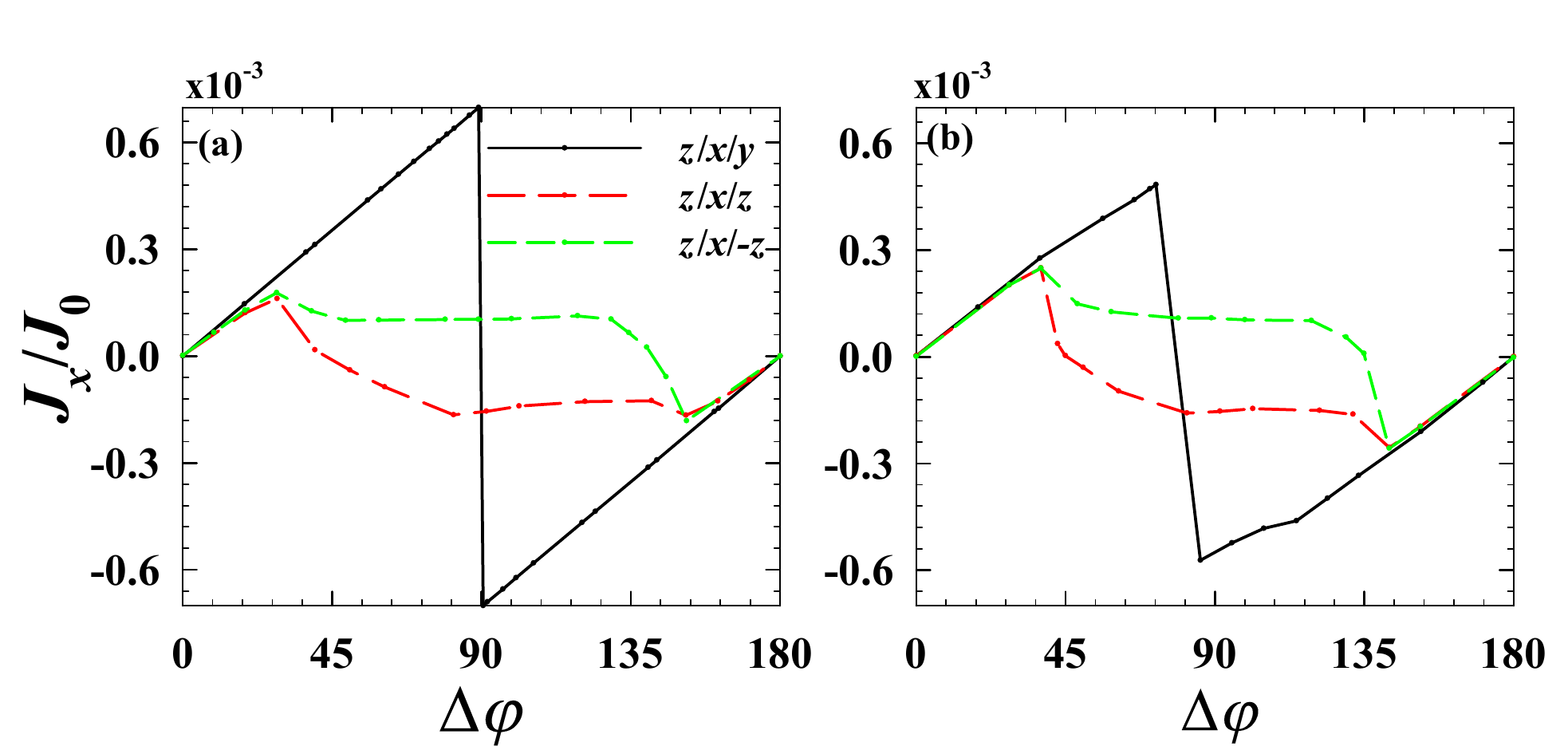}}
\caption{Normalized Josephson current 
versus $\Delta\varphi$ 
for a $SF_1 F_2 F_3 S$ structure. The legend labels
the three relative magnetization directions between the $F$ layers.
They correspond in order to the configurations labeled N P and AP
in the previous figures for $F_1$ and $F_3$, while the intermediate
layer $F_2$ has out-of-plane magnetization. 
The ferromagnetic layer $F_3$ has a fixed width corresponding to
$D_{F3}=380$.
The adjacent $F$ layers have widths 
(panel (a)) $D_{F1}=D_{F2}=10$, and (panel (b))
$D_{F1}=D_{F2}=20$.
 }   
\label{fig5}
\end{figure}
To examine the effects of increased magnetic inhomogeneity, 
we show in Fig.~\ref{fig5} results for a pentalayer $S F_1 F_2 F_3 S$
junction. This structure is complementary to that studied in
Fig.~\ref{fig2}, the main difference being an additional 
ferromagnet layer $F_2$ with an 
out-of-plane magnetic exchange field 
oriented in the $x$ direction 
(corresponding to  $\theta_2=0^\circ$ in
Fig.~\ref{fig0}). The relative magnetic orientations 
are labeled in the legend by the directions of the axes along
with the magnetizations are aligned in each $F$ layer. For example,
$z/x/y$ denotes a sample in which $F_1$ and $F_3$ are normal to
each other (the configuration labeled N in the previous figures)
with an additional out-of plane magnetization in $F_2$.
The width of the $F_2$ layer is identical to that of the $F_1$ layer:
$D_{F1}=D_{F2}=10$ in (a), and 
$D_{F1}=D_{F2}=20$ in (b). The thicker ferromagnet $F_3$ has
width $D_{F3}=380$. 
The $\pi$-periodic CPR that arises in double magnet $S F_1 F_2 S$ junctions
with orthogonal in-plane magnetizations
remains relatively unchanged by the addition of the additional 
out-of-plane intermediate ferromagnet.
However, it was shown in Figs.~\ref{fig2}(a)-(e)  that
the collinear P or AP magnetic states with still $2\pi$ periodicities 
behaved in an approximately piecewise  linear fashion and that the 
current maintained its direction
when varying $\Delta\varphi$. 
Now, on the other hand, the insertion
of an additional $F$ layer between the 
collinear ferromagnets, makes it possible for 
equal-spin triplet correlations to be generated and the supercurrent 
becomes drastically modified. For phase differences in the vicinity
of $0$ or $\pi$, 
Figs.~\ref{fig5}(a) and (b) show that the
current is  approximately linear in the phase difference,
but  there is a broad intermediate range of $\Delta \varphi$,
where the supercurrent flow is  relatively uniform.
Moreover, for either the P or AP configuration, 
varying the phase can result in the
Josephson current switching direction.
These trends are the same for each of the cases presented in (a) or (b).

\subsection {Magnetic orientation and CPR}
\label{mocpr}

\begin{figure}
\centerline{\includegraphics[width=8.8cm]{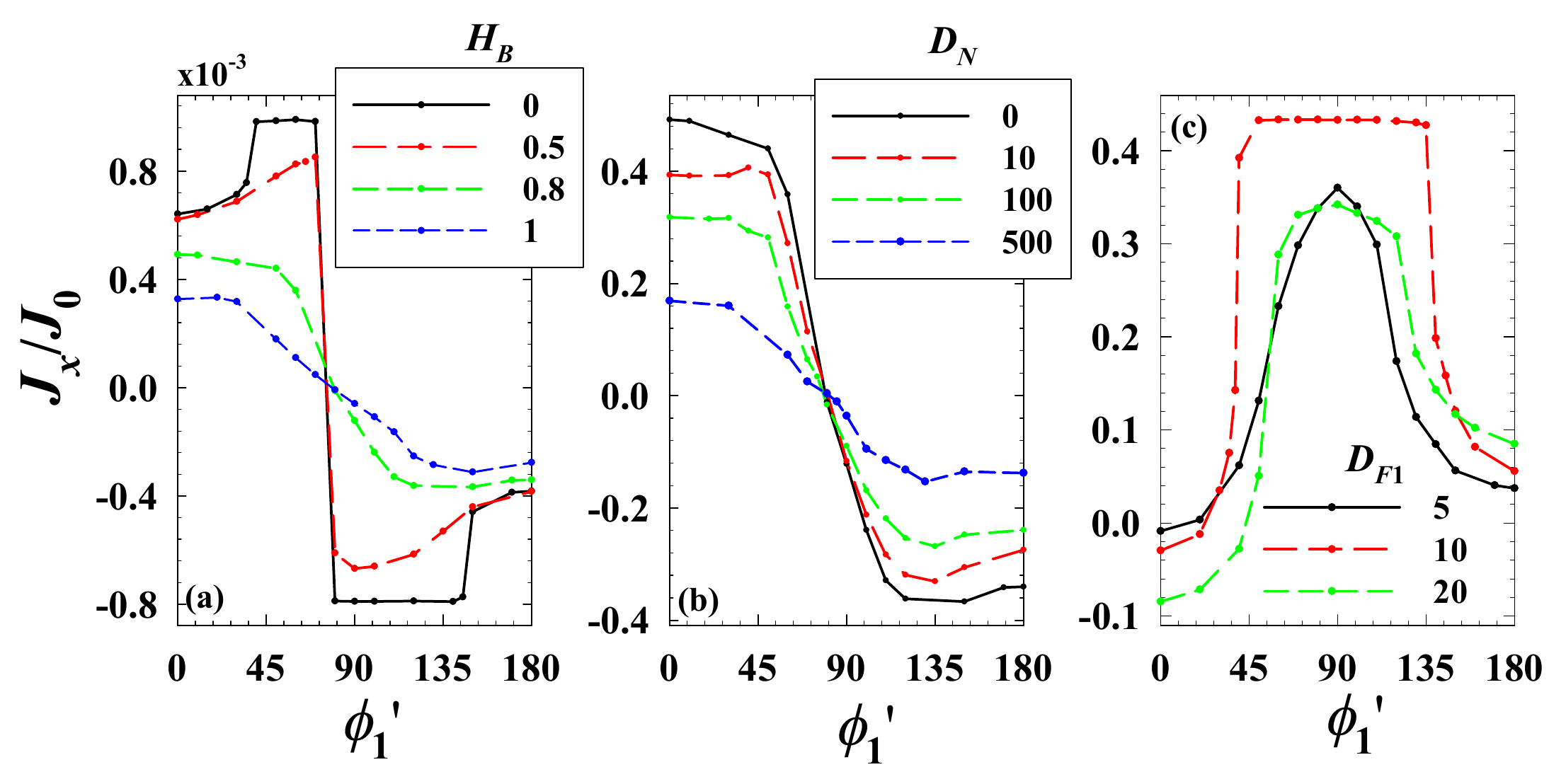}}
\caption{Normalized Josephson current 
versus the in-plane relative magnetization angle $\phi'_1\equiv \phi_1+
90^\circ$ (see text), 
for a $SF_1F_2S$ junction (panel (a)), and  
for an $SF_1NF_3S$ junction with $H_B=0.8$ (panel (b)). 
The legends show the 
geometric and material parameters that are varied.
In (a) and (b) $\Delta\varphi=100^\circ$, $D_{F1}=10$, and $D_{F2}=100$.
In (c) an $SF_1F_2S$ junction is considered with $H_B=0$ and 
$\Delta\varphi=60^\circ$.
The outer ferromagnet layer has $D_{F2}=380$, while 
three different $D_{F1}$ values are considered (see legend).
 }   
\label{fig6}
\end{figure}

Having discussed the current phase relations
for a few different ferromagnetic Josephson junction configurations,
we now will study in more detail the effect of varying magnetization
orientations on the CPR. We will
set the
macroscopic phase difference to a prescribed value
and study the supercurrent response
within the junctions for 
a range of relative magnetization
orientations.

Beginning with a basic $S F_1F_2S$ Josephson junction
whose phase difference is set at $\Delta\varphi=100^\circ$, 
we consider in Fig.~\ref{fig6}(a), the normalized supercurrent density
as a function of the in-plane magnetization angle $\phi'_1$ ($\theta_1=90^\circ$),
where  we introduce $\phi'_1\equiv\phi_1+90^\circ$, with
$\phi_1$ being the angle shown in Fig.~\ref{fig0}. 
In terms of $\phi'_1$,
the P and AP states then correspond to 
$\phi'_1=0^\circ$, and $\phi'_1=180^\circ$ respectively
(since the magnetization in $F_2$ is  fixed 
along $y$).
Four interface scattering strengths are considered in panel (a), as 
indicated in its legend.
In all cases, by tuning the relative alignment angles,
supercurrent switching occurs when the mutual magnetization orientations
are approximately orthogonal. 
As expected, the
supercurrent flow is greatest for transparent interfaces ($H_B=0$),
and decreases with increasing $H_B$, as the sensitivity 
to $\phi'_1$ becomes weaker.
It is notable that the maximum current flow occurs at different 
$\phi'_1$ values,
depending on the  interface scattering strength. 
In Fig.~\ref{fig6}(b), an additional $N$ layer, of variable
strength as indicated,  is inserted between
the two ferromagnets and the interface 
scattering   parameter is set to $H_B=0.8$.
The solid $D_N=0$ curve corresponds to the $H_B=0.8$ curve in panel (a).
The figure shows that
increasing the $N$ layer thickness
tends to generally dampen the current
through the junction.
The
current flow peaks when the two 
ferromagnets are aligned in the P state  
and it 
vanishes altogether near the N configuration, 
before reversing direction as
the relative magnetizations approach the AP state. 
In Fig.~\ref{fig6}(c) we investigate
the supercurrent flow in a wide $S F_1 F_2 S$ junction, with $D_{F2}=500$, 
and three different $F_1$ widths (see legend).
The macroscopic phase difference is set at $\Delta\varphi=60^\circ$.
It is seen that for this geometry,
the current has some contrasting features compared to the previous cases
involving thinner $F$ layers.
For instance, when tuning $\phi'_1$,
now
the weakest current flow occurs when in the P state,
and the maximum occurs in the orthogonal configuration. 
For all $F$ width considered, the current undergoes 
rapid changes in the vicinity of the middle value between N and P or N and
AP configurations, 
with the $D_{F1}=10$ case changing the most,
and then remains nearly constant at angles near the P and AP
configurations.
Since the prescribed phase difference
$\Delta\phi=60^{\circ}$ is near where the maximum currents occur
in $\pi$-periodic Josephson junctions,
it is evident that the results in  Fig.~\ref{fig6}(c) are a direct consequence 
of the induced equal-spin triplet correlations.
We will see below in Sec.~\ref{itp}
that this correlates with the triplet generation behavior. 

\begin{figure}
\centerline{\includegraphics[width=8.8cm]{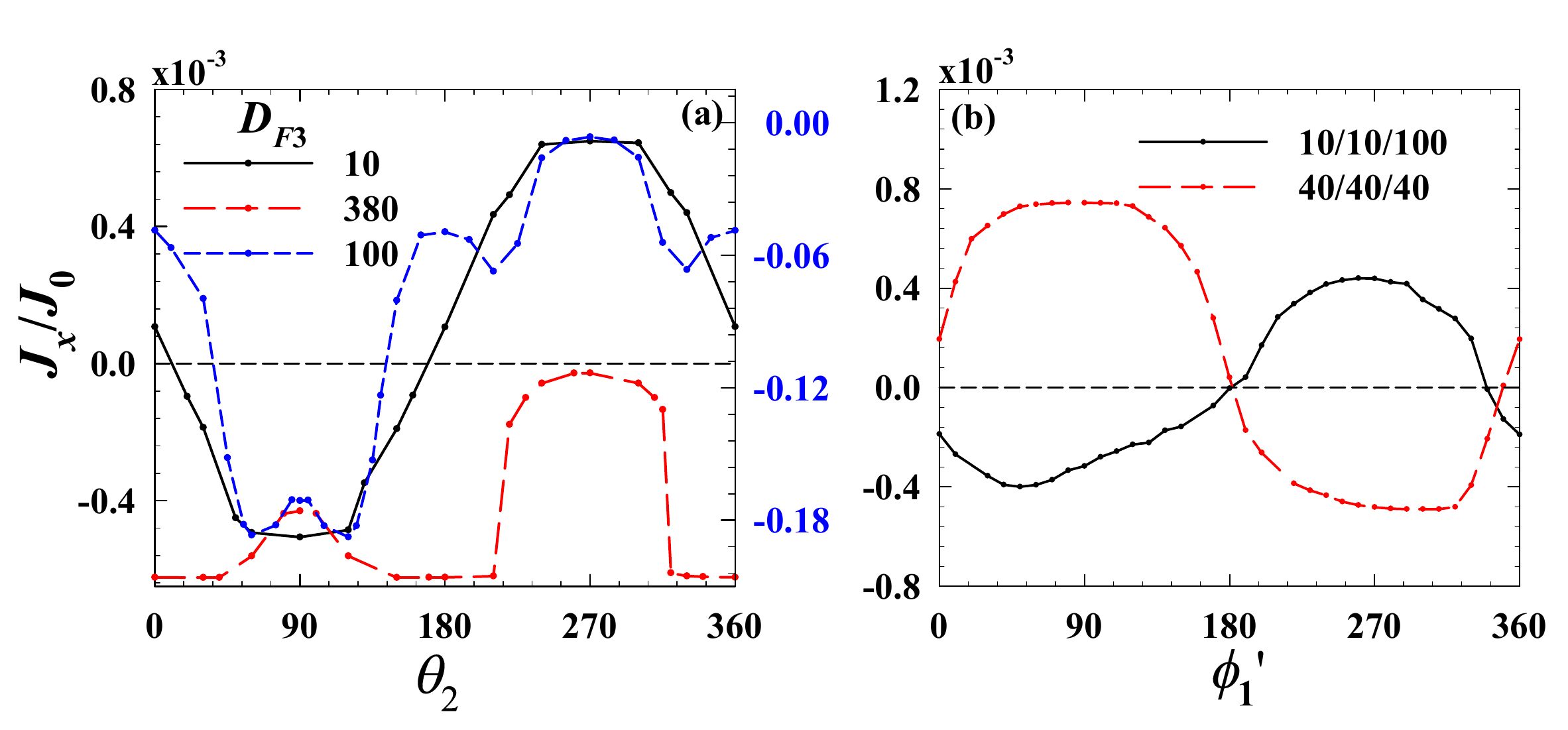}}
\caption{ Normalized Josephson current for a $S F_1 F_2 F_3 S$ junction  
versus  (panel (a)) the out-of-plane angle 
$\theta_2$ (see Fig.~\ref{fig0}), and (panel (b))
 the in-plane angle $\phi'_1$.
A set phase difference  $\Delta \varphi=100^\circ$
 is assumed for all cases.
In (a) the ferromagnets, $F_1$ and $F_2$ have widths 
$D_{F1}=D_{F2}=10$, and the legend lists the three $F_3$ widths considered.
The exchange field points along the $y$ direction in $F_3$ and the 
$z$ direction in $F_1$.  
The right vertical axis is for the $D_{F3}=100$ case only. 
For the $D_{F3}=380$ case, there is no interface scattering ($H_B=0$), 
while the remaining $D_{F3}$ cases
have $H_B=0.8$.
In (b) $\theta_1=90^\circ$, and $\bm{h}$ is 
directed along $y$ in $F_2$, and along $z$ in $F_3$.  
The legend indicates the $F$ layer
widths  of the two trilayer ferromagnet structures considered:
they have the same  total width.
Interface scattering in both cases is $H_B=0.8$. 
 }   
\label{fig7}
\end{figure}
We next move to the study of 
the magnetization orientation role in the supercurrent for
more complicated $S F_1 F_2 F_3 S$ junctions.
As an example, we consider a scenario where 
the $F_1$ and $F_3$ layers have
magnetizations pinned along  the $z$ and $y$  
directions respectively,
that is, a relative  N configuration, 
but the magnetization in the central $F$ layer rotates 
on the $xz$ plane ($\phi_2=0$) from along the $x$-axis to
 the $z$-axis. 
In  
Fig.~\ref{fig7}(a), we show the normalized current density 
as a function of $\theta_2$ for few different $F_3$ layer widths, $D_{F3}$.
We set $D_{F1}=D_{F2}=10$,
and consider values of
$D_{F3}$ that lead to both symmetric and asymmetric structures.
In each case,
$\Delta\varphi=100^\circ$, and
interface scattering is present with $H_B=0.8$,
except for the widest junction  with $D_{F3}=380$
where the interfaces are transparent ($H_B=0$). 
In the case $D_{F1}=D_{F2}=D_{F3}=10$,
where all $F$ thicknesses are identical, 
the current flow is approximately antisymmetric around 
$\theta_2=180^\circ$. 
The current flow is suppressed 
for the asymmetric $D_{F3}=100$ 
situation by  decoherence effects
arising from the larger width.
While $J_x$ for the simpler $S F_1 F_2 S$ junctions in
 Fig.~\ref{fig6} was $\pi$ periodic in $\phi'_1$,
variations in $\theta_2$ 
for trilayer ferromagnetic junctions
are in general $2\pi$ periodic, as seen in the figure.
The asymmetric case also demonstrates a more intricate structure 
as the magnetization angle is swept.
For the narrower junction with $D_{F3}=10$, the orientations 
in which the current
switches direction are near $\theta_2=0^\circ$ and $\theta_2=180^\circ$.
When $D_{F3}=100$, the current never changes its direction.
Although the rotating out-of-plane exchange field of $F_2$
does affect the strength of current with $2\pi$ periodicity.
For the highly asymmetric $D_{F3}=380$ case, 
the current again maintains its direction of flow over the full angular range of $\theta_2$.
Also, the current is approximately  constant except for 
 orientations when the exchange field in $F_2$ points along $z$: 
 $\theta_2=90^\circ$ or  $\theta_2=270^\circ$.
 For orientations near $\theta_2=270^\circ$,  
 we find that the current is strongly suppressed. 
In 
Fig.~\ref{fig7}(b),
the 
in-plane 
magnetization in 
the first ferromagnetic layer is  
now  varied,  
while the other two are kept fixed: for $F_2$, it is along $y$, and
for $F_3$, it is along $z$, an N configuration.
Two types of structures are considered, with the total 
width of the three ferromagnetic regions being constant. 
In the first case, $D_{F1}=D_{F2}=10$, and $D_{F3}=100$,
while in the second one all three $F$ layers have width, 
$D_{F1}=D_{F2}=D_{F3}=40$.
Both cases exhibit similar behavior as a function of $\phi'_1$. 
The current vanishes when $F_1$ 
is antiparallel to  $F_2$ 
and is highest when the magnetization lies nearly in between the
those of $F_2$ and $F_3$. 
Thus, the charge supercurrent which flows
oppositely in the two structures, can be effectively 
switched on or off by manipulating the in-plane magnetization angle of the
first ferromagnet.

\begin{figure}
\centerline{\includegraphics[width=8.0cm]{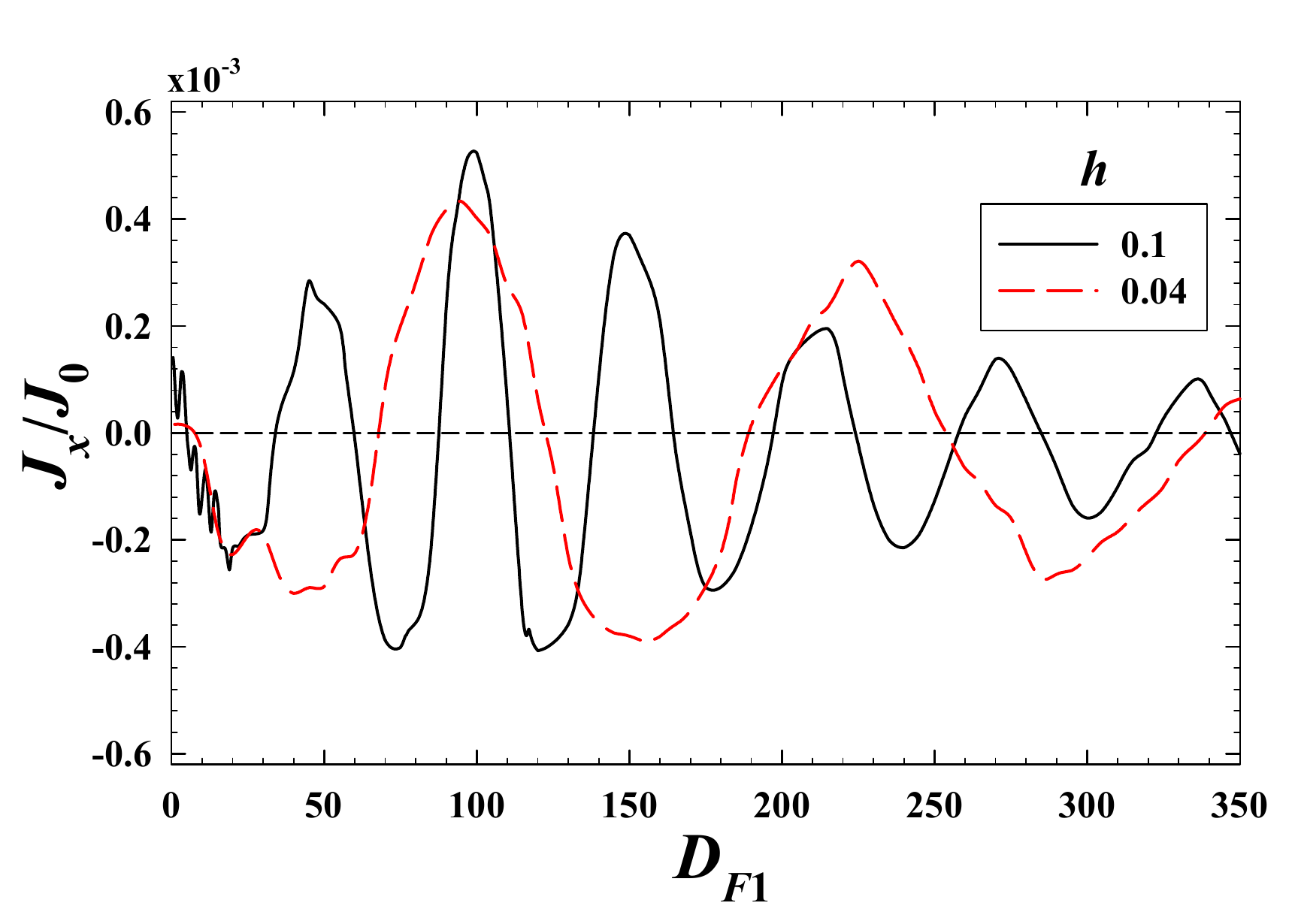}}
\caption{Normalized Josephson current for a $S F_1 F_2 S$ structure as a function of  $F$ layer width $D_{F1}$.
The other ferromagnet's width is fixed at $D_{F2}=100$.
The exchange field is along $z$ in $F_1$, and along $y$ in $F_2$.
The legend indicates two normalized $h$ values: 
The dashed curve corresponds to  $h=0.04$, 
while the solid curve is for $h=0.1$ 
The interface scattering parameter is set to $H_B=0.8$,
and a phase difference of $\Delta\varphi=100^\circ$ is maintained across the $S$ electrodes.
 }   
\label{fig8}
\end{figure}
It is known that in $F/S$ heterostructures, including bilayers\cite{klaus} and
Josephson junctions, variations in the magnetic exchange field and 
ferromagnet thickness
induce damped oscillations
in the spatial behavior of the Cooper pair amplitudes, 
resulting in modulation of  physical quantities as a function 
of either $h$ or $D_F$.
The damped oscillations in the clean limit have a  wavelength 
that goes as the
inverse of the exchange field.
To investigate this phenomenon 
in an $S F_1 F_2 S$ junction, we show in Fig.~\ref{fig8},
the supercurrent that flows through the junction as a function of the 
$F_1$ width,
$D_{F1}$. 
For the adjacent $F_2$ layer, the  width is 
$D_{F2}=100$,
and with interface scattering strength $H_B=0.8$. 
Two exchange fields are considered: 
$h=0.04$ and $h=0.1$. 
The period of oscillations in each case 
is seen to be 
approximately $2\pi k_F \xi_F$, where $k_F \xi_F\equiv \varepsilon_F/h = 25$,  and $10$,
respectively.
The current for both cases is maximal when the two $F$
regions are equal, and then slowly dampens out with increasing $D_{F1}$.
This decay length is inversely proportional to the magnitude of the exchange field.
Moreover, the charge current 
in each case periodically changes sign.
Since the oscillations in the current as a function of $D_{F1}$
increase with larger exchange fields, the current direction can be very sensitive to 
fabrication tolerances for strong magnets.

\begin{figure}
\centerline{\includegraphics[width=8.8cm]{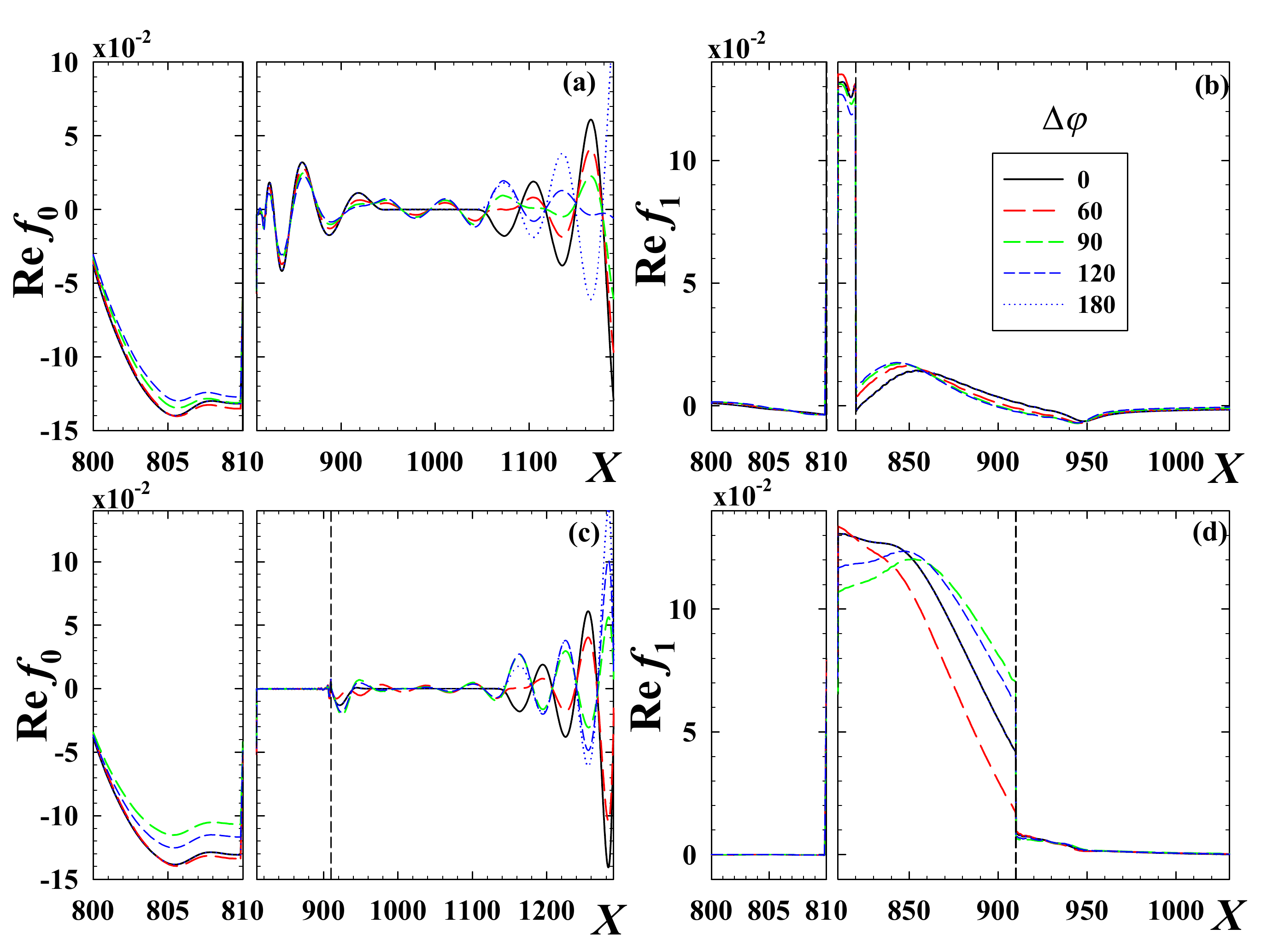}}
\caption{Normalized triplet correlations in an $SF_1NF_3S$ Josephson junction
as a function of position.  The $F$ layers have widths $D_{F1}=10$ 
and $D_{F3}=380$,
and $H_B=0$, corresponding to the parameters used in Fig.~\ref{fig4}(c).
The top set of panels relate to structures with a normal metal spacer, $N$, 
of width $D_N=10$,
while the bottom set represent a larger $N$ layer with $D_N=100$. 
The dashed  
vertical lines represent the
interface between the $N$ and $F_2$ regions.
Various 
phase differences $\Delta \varphi$ are considered (see legend).
The magnetization in $F_1$ is along $z$, while it is along $y$ in $F_2$.
 }   
\label{fig9}
\end{figure}

\subsection{Induced triplet pairing}
\label{itp}
We now discuss the 
induced triplet pairing correlations in ferromagnetic Josephson junctions.
The presence of  multiple misaligned ferromagnets
yields  both the $m = 0$ (Eq.~(\ref{f0})) and the $m=\pm1$ (Eq.~(\ref{f1}))
triplet pair amplitudes as permitted  by conservation laws
and the Pauli principle.
To gain an overall view of the opposite-spin triplet amplitudes, $f_0$,
and the equal-spin amplitudes, $f_1$,
we illustrate in Fig.~\ref{fig9} the spatial behavior of these correlations 
in the $N$ and $F$ regions of an
$S F_1 N F_2 S$ junction.
We focus on the real parts of  these  complex quantities,
keeping in mind that the imaginary components obey similar trends.
The geometrical parameters in this
figure are $D_{F1}=10$, $D_{F2}=380$,
and $D_N=10$ (top panels) or $D_N=100$ (bottom panels).
Thus, in all  panels the region $800<X<810$
is occupied by $F_1$ while $F_2$ occupies  the region $820<X<1200$ in
the top panels and $910<X<1290$ in the bottom panels (c) and (d),
where the vertical dotted line denotes the $N$ spacer boundary.
Hence,
different horizontal scales are used in each case.
The scattering parameter is set to $H_B=0$, and
each curve corresponds to a different phase difference, $\Delta\varphi$
as indicated by the legend. The exchange fields are
in-plane and normal to each other.
Within the $F_1$ region, panels (a) and (c) 
reveal that the magnitude of the $f_0$  pair correlations
are approximately of the same magnitude,  decreasing in the 
vicinity of the $S/F$
interface located at $X=800$.
The system with the wider normal metal layer, $D_N=100$ is  slightly more
sensitive to phase variations. 
Within the ferromagnet $F_2$, the same panels (a) (c) show the
oscillatory nature of $f_0$, which behaves similarly to the singlet 
pair amplitude, the periodicity
arising from the difference in spin-up and spin-down wavevectors.
For the chosen exchange field, 
the oscillations are limited in $F_1$ due to the confined width.
Turning now to the equal-spin triplet correlations $f_1$, 
panels (b) and (d) display
behavior which contrasts with the $f_0$ results. In particular, 
within the narrow $F_1$ region,
the $f_1$ triplets are negligibly small, and the $f_0$ correlations 
clearly dominate.
In the $N$ region,  the $f_0$ correlations nearly vanish, 
while the equal-spin triplets
peak near the $F_1/N$ interface
(at $X=810$), before dropping 
 within the normal metal (see Fig.~\ref{fig9}(d)). 
Finally, within $F_2$, the triplets $f_1$ assume
a slow, long-range variation compared to the damped oscillatory 
behavior of the $f_0$ curves.

\begin{figure}
\centerline{\includegraphics[width=8.8cm]{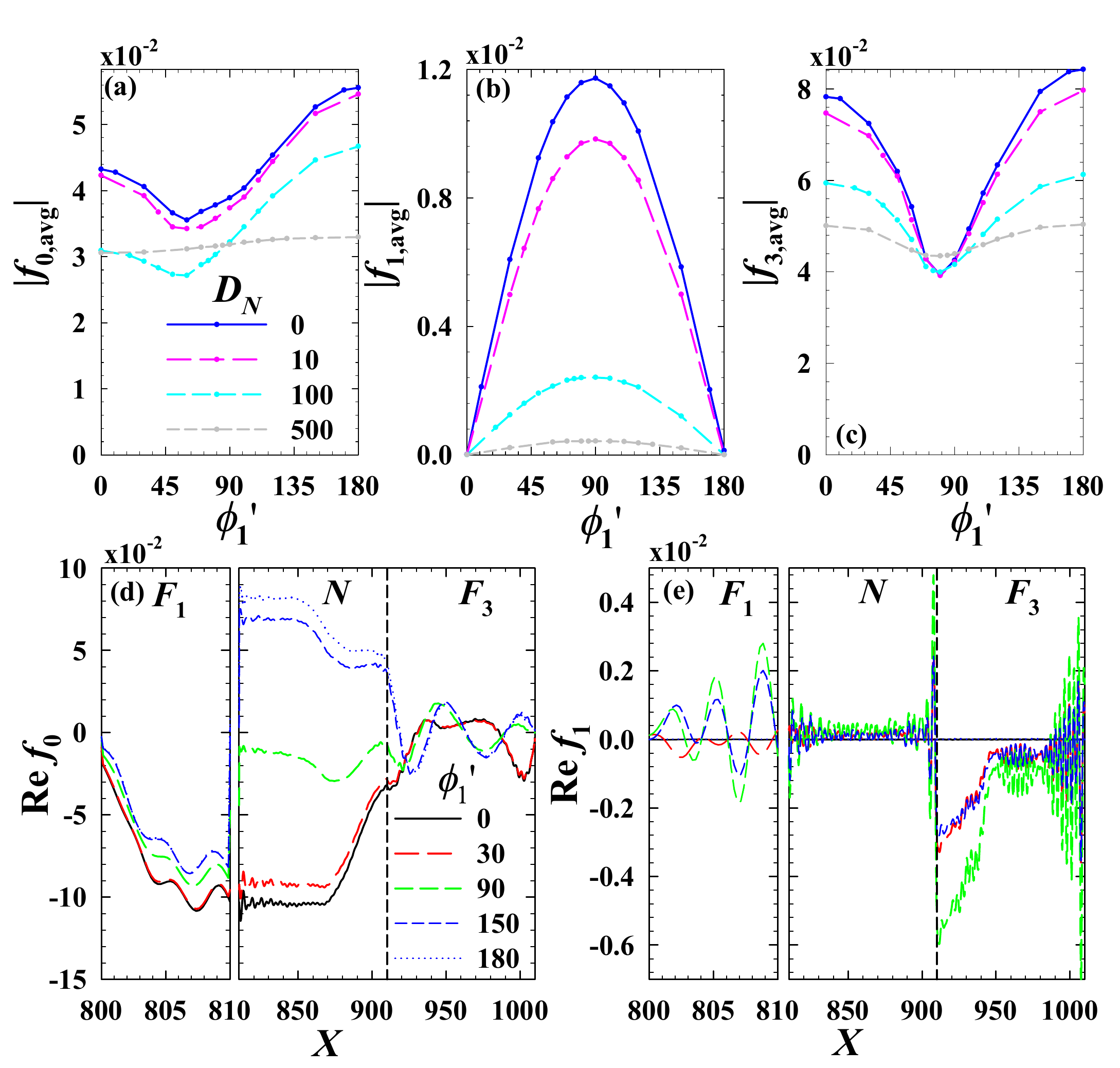}}
\caption{Top row, panels (a)-(c): Normalized  singlet and triplet correlations
versus  in-plane relative magnetization angle 
$\phi'_1\equiv \phi_1+90^\circ$ for a 
$S F_1 N F_2 S$ structure.
The magnitudes of these pair correlations are averaged over the $F_2$ region. 
The geometrical and material parameters are the same as  in 
Fig.~\ref{fig6}(b).
The bottom panels (d) and (e) correspond to the local spatial behavior of the  
triplet correlations for the $D_N=100$ case
studied in the top panels. The two $F$ regions are plotted in 
separate frames to
discern the triplet correlations in the narrow $F_1$ region. 
The dashed 
vertical lines represent the
interface between the $N$ and $F_2$ regions.
Several  values of $\phi'_1$ are considered, 
as shown in the legend. 
}   
\label{fig10}
\end{figure}
In addition to investigating the spatial
behavior of the triplet amplitudes,
 it is instructive to also examine  the spatially averaged triplet 
 and singlet correlations as
functions of the relevant system parameters.
For example, by tuning
 the relative magnetization angle, (varying $\phi'_1$ at 
 fixed $\phi_2=90^\circ$), 
important  overall features can be revealed.
The top panels (a) (b) (c) of
Fig.~\ref{fig10}  show  the $\phi'_1$-dependence of the 
magnitudes of the triplet and singlet
amplitudes averaged over the $F_2$ region 
for the $S F_1 N F_2 S$ structure studied initially in Fig.~\ref{fig6}(b).
Four representative $N$ layer widths are considered
as indicated in the legend. 
The proximity effects and hence coupling of the two ferromagnets
diminish with increasing $D_N$, and therefore 
the pair correlations become less sensitive to
variations in $\phi'_1$, as observed for the largest $D_N=500$ case.
Other than the diminished magnitudes, the overall trends and behavior however
do not depend strongly on the presence of the normal metal spacer.
This may be   important in experiment design, where spacers 
are often needed.
The opposite-spin triplet correlations, $f_0$ and 
the singlet pair amplitude ($f_3\equiv \Delta/g$) 
in (a) and (c)  
behave in rather similar ways,
but  $f_3$  is more symmetric about the orthogonal direction 
$\phi'_1=90^\circ$. 
When the relative magnetization orientation
 varies in inhomogeneous $S/F$ systems,
the process of singlet-triplet conversion plays a role in the transport 
and thermodynamic properties of such systems.
It is apparent from panels (a) 
and (b) that the 
orientation 
$\phi'_1$ that 
leads to a
minimum in $f_0$ and $f_3$, 
corresponds to that where  $f_1$
 is largest. 
These occurrences arise when 
the exchange fields in $F_1$ and $F_2$ are nearly orthogonal. This angle
for $f_0$ slightly shifts with variations in $D_N$. 

\begin{figure}
\centerline{\includegraphics[width=8.8cm]{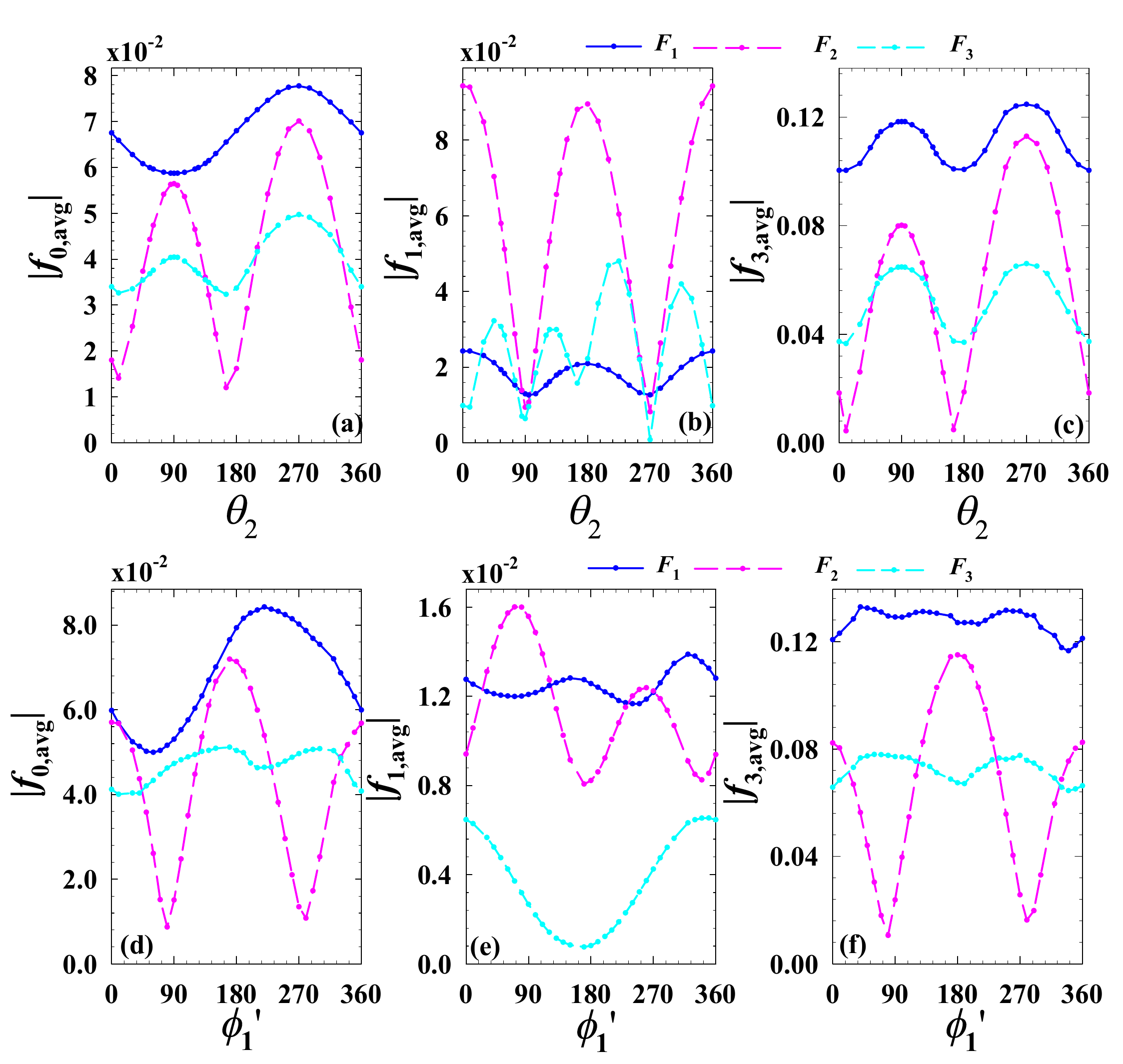}}
\caption{
Normalized triplet $|f_{0,{\rm avg}}|$, $|f_{1,{\rm avg}}|$, and singlet  
$|f_{3,{\rm avg}}|$ amplitudes, averaged over the $F$ regions indicated
on the overhead legends, plotted 
as functions of $\theta_2$ and $\phi'_1$ 
The geometrical and material parameters corresponds to the  $D_{F2}=100$
cases  presented in Fig.~\ref{fig7}.
}   
\label{fig11}
\end{figure}

In the bottom panels (d) and (e) of Fig.~\ref{fig10}  
we display the 
spatial dependence of the real components of the 
triplet correlations throughout each of the three junction regions
discussed in the top panels.
Results for five different relative magnetic orientations are presented, 
(see legend) including the $P$ ($\phi'_1=0^\circ$),
AP ($\phi'_1=180^\circ$), and N ($\phi'_1=90^\circ$) 
configurations. 
Considering first Fig.~\ref{fig10}(d) 
in the $F_1$ region, we see behavior 
 similar to that found for the wider 
junction case in Fig.~\ref{fig9}, including a relatively weak dependence on the
orientation angle $\phi'_1$ (as opposed to the phase difference $\Delta\varphi$).
The central $N$ region is most affected by variations in  $\phi'_1$: 
the real part of $f_0$ 
changes sign
when $\phi'_1$ is swept 
from the P to AP state, and nearly vanishes altogether at 
$\phi'_1\approx 90^\circ$. In $F_2$, a series of oscillations emerge with a 
periodicity similar to that
found in Fig.~\ref{fig9}, since an exchange field of the same strength was used.
In Fig.~\ref{fig10}(e), the equal-spin $f_1$ amplitudes exhibit considerably 
different
behavior. First, only three of the considered $\phi'_1$ yield nonzero
results, 
since the P
and AP configurations cannot generate equal-spin triplet correlations.
Interestingly, within the $F_1$ region  $f_1$  does not exhibit a slow decay, 
but rather oscillates
with a period that is much shorter than the opposite spin pairs  governed by 
the difference in
spin-up and spin-down wave vectors.
We find that within the normal metal layer, there is nearly a complete 
absence of equal-spin correlations, this is
accompanied by the appearance of  opposite-spin correlations $f_0$ 
(see panel (d)).
The $f_1$ amplitudes are largest in the $F_2$ layer,  for the relative
orientation of $\phi'_1=90^\circ$, in agreement with the averaged results 
in Fig.~\ref{fig10}(b).
For the relative magnetization angles of $\phi'_1=30^\circ$,  
and $\phi'_1=150^\circ$,
the $f_1$' amplitudes are identical due to the symmetry about $\phi'_1=90^\circ$.
One can now correlate the features Fig.~\ref{fig10} and \ref{fig11} to Fig.~\ref{fig4}.
From Fig.~\ref{fig4}, we learn that the N magnetic configuration often leads
to the appearance of $\pi$-Josephson junctions. Here we are able to 
give concrete proof that in the N cases the equal-spin 
triplet correlations are maximized and are insensitive to 
$\Delta\varphi$.  
Therefore, the CPRs for different magnetic configurations are 
essentially characterized by their detailed singlet/triplet nature.

We next examine, in Fig.~\ref{fig11},
the behavior of the averaged triplet and singlet amplitudes,
as the magnetic orientation angles, $\theta_2$ (top panels), and $\phi'_1$
(bottom panels) are changed,
in more complicated $S F_1 F_2 F_3 S$ Josephson junctions. 
This study is therefore complementary to the results shown 
in Fig.~\ref{fig7} involving
the charge supercurrent.
The geometric parameters are $D_{F1}=D_{F2}=10$, and $D_{F3}=100$.
The region in which the pair 
correlations are averaged over is 
specified in the top legends.
In the top row of Fig.~\ref{fig11} 
we present results  for
magnetization orientations $\theta_2$ sweeping the
entire angular range from $0^\circ$ to $360^\circ$, while
the magnetizations are aligned along $z$ in  $F_1$  and along $y$ in
$F_3$. Therefore when $\theta_2=0^\circ$ or $\theta_2=180^\circ$,
all three ferromagnets have mutually orthogonal magnetizations, 
corresponding to a high degree of magnetic inhomogeneity.
Under these circumstances, one can expect
that the equal-spin triplet correlations $f_1$
should be, on the average,  at their highest values, while the opposite spin 
correlations should be
weakest.
Indeed, in the regions $F_3$ and
$F_2$, the opposite-spin singlet $f_3$, 
and triplet $f_0$ correlations
possess minima near these angles. 
In contrast to the spatially
averaged $f_1$ amplitudes  peak, in $F_3$ and $F_2$, at those orientations. 
Although the general trends 
are usually the same for all, $F$ layers,
proximity geometrical effects can result in
self-consistent triplet correlations with 
more intricate nontrivial structure, and  this is the case with 
the averages over the $F_1$ region. 
In the bottom 
set of panels of  
Fig.~\ref{fig11},
we consider in-plane magnetization rotations of the $F_1$ layer.
The other ferromagnets $F_2$ and $F_3$ have their magnetizations fixed
in the $y$ and $z$ directions respectively. 
For this situation, the opposite-spin $f_0$ amplitudes in $F_1$
are seen to be $2\pi$ periodic, peaking
at $\phi'_1 \approx 225^\circ$. In $F_2$ these triplet amplitudes
are seen
to be largest when the relative orientations between $F_1$ and $F_2$
are either P ($\phi'_1=0^\circ,360^\circ$) or AP ($\phi'_1=180^\circ$).
This is consistent with the behavior of the triplet amplitudes
found in double magnet spin valve systems \cite{wvh14}. 

\subsection{Spin Transport}
\label{st}

\begin{figure}
\centerline{\includegraphics[width=8.8cm]{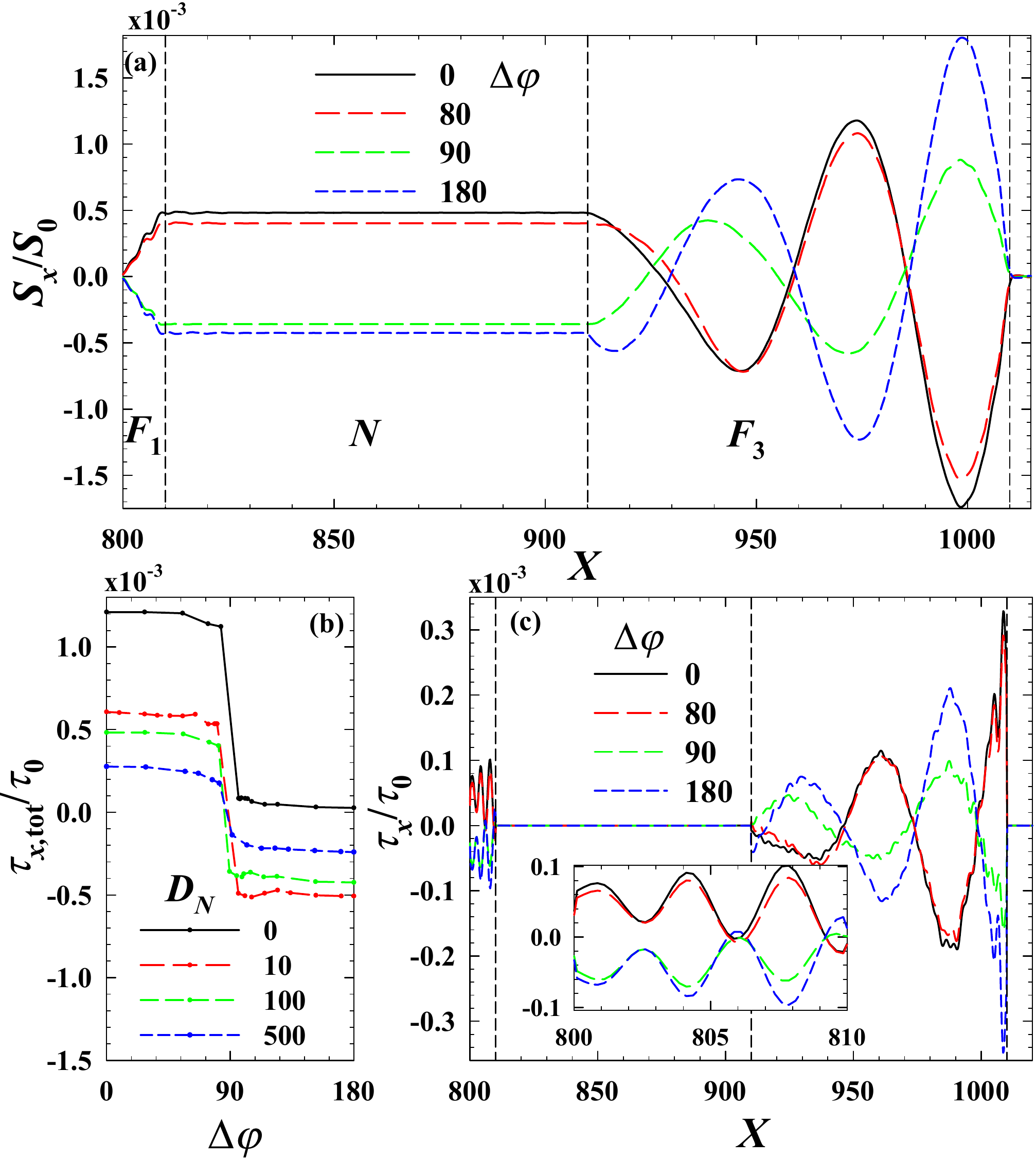}}
\caption{Panel (a): Normalized 
$x$ component of the spin current in an $SF_1 N F_2 S$ Josephson junction
vs. position.  
The geometric widths are $D_{F1}=10$, $D_{F2}=100$,
and $D_N=100$. 
There is moderate interface scattering with $H_B=0.5$ 
(see e.g., Fig.~\ref{fig4}(a)). 
The dashed vertical lines
mark the interfaces.
Panel (b) The total torque $\tau_{x,{\rm tot}}$ 
acting within the $F_1$ region as a function of  
$\Delta\varphi$.
Several normal metal widths are considered, see
legend. The $D_N=0$ case
has been shifted downwards by $4\times 10^{-3}$ 
for comparison purposes. 
In panel (c) the $x$-component of 
the local torque is shown vs. position with the same phase
differences used in (a).
The inset is a magnification of the torque within 
the  narrow $F_1$ region ($800<X<810$).
 }   
\label{fig12}
\end{figure}

Having established the salient features of 
supercurrent charge transport and pair correlations 
in a variety of
ferromagnetic Josephson junctions,
we now explore 
 the spin degree of freedom
and determine
the crucial spin currents
and the associated spin transfer torques. 
The current that is 
generated from the macroscopic phase differences between the $S$
electrodes
can become spin-polarized\cite{brou,Shomali} 
when entering one of the ferromagnetic
regions.
This spin current can then interact with
the other ferromagnets and be modified by the local  
magnetizations due to the spin-exchange
interaction, via the existence of spin transfer torques. 
The conservation law associated with process is described by 
Eqs.~(\ref{tau1}) and (\ref{sx1}). 
It is  important  not only to understand the behavior of the
spin-polarized currents in  ferromagnetic Josephson junctions, but
also the various ways in which to manipulate them
for practical
spintronic applications.
We therefore investigate
from a  microscopic and self-consistent perspective,
the equilibrium spin
currents and associated 
torques throughout the entire junction regions
as 
functions of position, phase difference, and magnetization orientation
angles.

\begin{figure}
\centerline{\includegraphics[width=8.8cm]{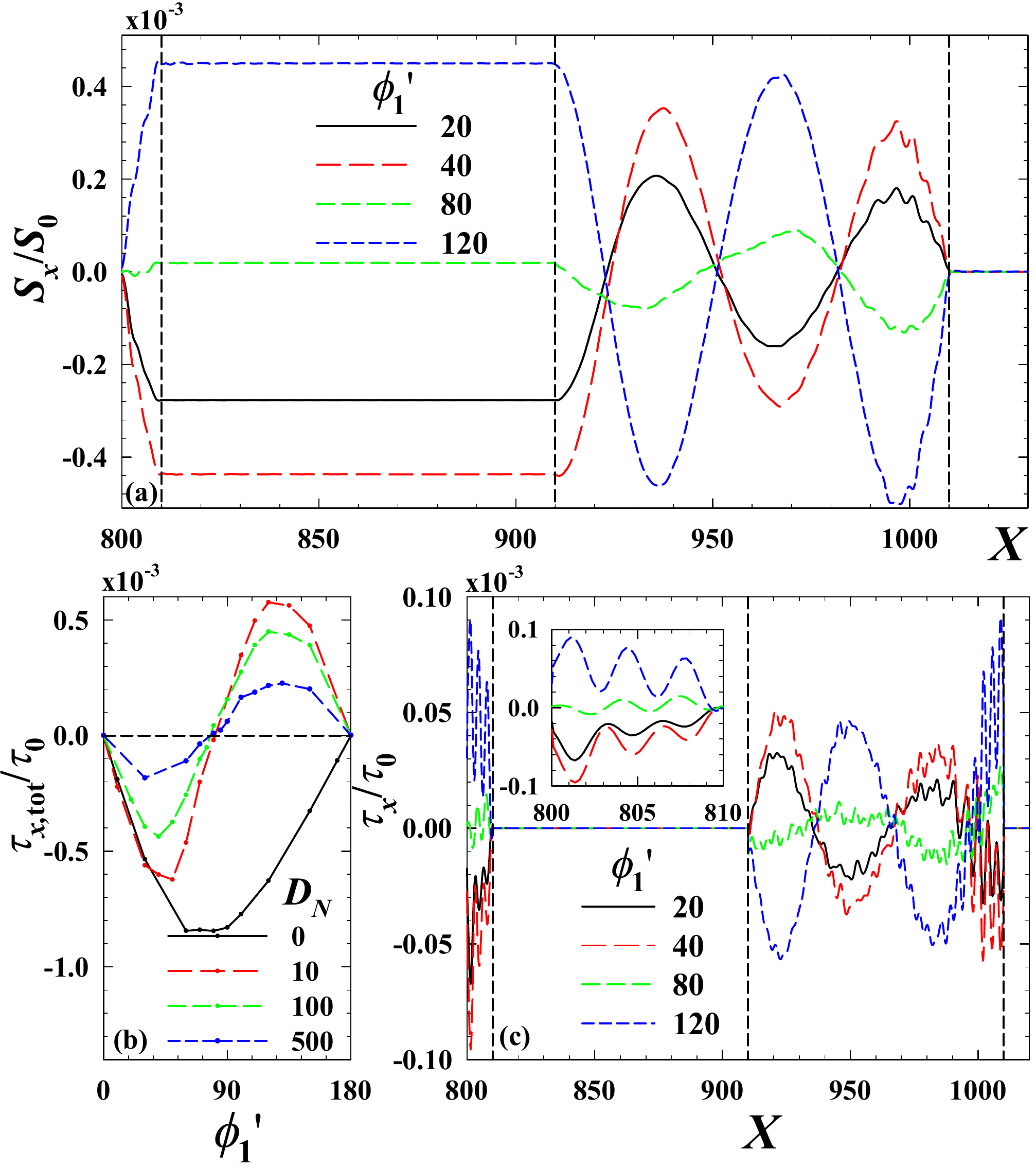}}
\caption{Panel (a): Normalized spin current in an $SF_1NF_2S$ Josephson junction
as a function of position.  The geometric widths are $D_{F1}=10$, 
$D_{F2}=100$,
and $D_N=100$. The interface scattering strength is set to $H_B=0.8$.
These values are the same as in Fig.~\ref{fig6}(b). The dashed vertical lines
mark the interfaces. Panel
(b) illustrates the total torque $\tau_{x,{\rm tot}}$ 
within the $F_1$ region as a function of the relative in-plane 
magnetization angle, $\phi'_1$.
Several normal metal widths are considered as depicted in the legend. 
The $D_N=0$ results
have been multiplied by $1/5$  for comparison purposes. In 
panel (c) the $x$-component of the torque is shown
as a function of position for the same case considered
in (a).
 }   
\label{fig13}
\end{figure}

In Fig.~\ref{fig12}(a), we consider the spatial dependence of the
spin current in a $S F_1 N F_2 S$ junction. 
The geometrical parameters used in this plot are 
the same as in Fig.~\ref{fig4}(a),
with $D_N=100$. Our geometry ensures that the
generally tensorial  spin current is reduced to 
a vector in spin space, representing a 
spin vector
current flowing  in the spatial $x$ direction and having in general three
components in spin space.  We
display the spatial dependence of 
the $x$ spin component, $S_x$, (normalized
as previously discussed). 
Because the exchange interaction (and hence the torque) 
vanishes in the  $N$ and $S$ 
regions, only  
the $F$ regions of the junction can 
have a spatially varying spin current: in the $N$ and $S$ regions
the spin current must be spatially invariant. 
Under our constant-phase and zero voltage boundary conditions 
the outer $s$-wave superconducting 
regions do not\cite{wvh14}  
support a spin current,
and hence ${\bm S}$ 
vanishes there. The central nonmagnetic $N$ layer, however,
couples the two ferromagnets via a constant spin current, which is
related to the net torque acting within the $F$ regions (see Eq.~(\ref{sx2})).
The spin current oscillates in the $F_2$ region.  The  
amplitude of these oscillations depends on
$\Delta\varphi$, while  the period does not: 
the points in $F_2$ where $S_x$ vanish are independent of $\Delta\varphi$.
If a ferromagnet is very thin,
as occurs for $F_1$, the spin currents vary nearly linearly with $X$,
which can be viewed as a small segment of a sinusoidal function.
To  present an overall view of how the change in spin current 
and its associated  torque vary 
as the phase varies. We do this in terms of the 
total torque, defined as the integral of the
local torque, normalized
as discussed above, over  dimensionless 
distance.  We plot in Fig.~\ref{fig13}(b)
the total torque, $\tau_{x,{\rm tot}}$, within $F_1$.  for a few  values of the 
interlayer $N$ spacer thickness.
In all cases, $\tau_{x,{\rm tot}}$ is relatively uniform until a
sharp crossover near 
$\Delta\varphi=90^\circ$, where the net torque changes sign,
coinciding with the point of  supercurrent reversal (Fig.~\ref{fig4}(a)).
Interestingly, only when the normal metal insert  is present, 
does $\tau_{x,{\rm tot}}$
reverse direction. 
Panel (c) illustrates the local $x$-component of the torque, $\tau_x$, as
a function of position throughout the entire junction region.
The inset is the same quantity, but plotted only over the narrow $F_1$ region.
To correlate with (a), the $D_N=100$ case is considered here.
Each curve represents a different phase difference as shown in the legend
for (a).
For in-plane exchange field  interactions, no other component of the torque 
can exist in equilibrium when
spin currents do not enter or leave the superconducting electrodes.\cite{brou}
Thus,
the net torque for the entire system must vanish,
requiring $\tau_{x,{\rm tot}}$ for each of the two $F$ regions 
to be opposite in sign,
despite the dissimilar spatial behavior  as exhibited in (c).
Comparing (a) and (c), it is also evident that within the oscillatory 
$F_2$ region, 
$\tau_x$ and $S_x$ behave similarly, but are out of phase
by approximately $90^\circ$, in agreement with 
Eq.~(\ref{sx1}).
Since the spin current was shown to be uniform in the normal metal region,
the torque is seen to vanish there, 
as it should be in regions where the magnetic 
exchange interaction is absent.

We now turn to studying how the spin currents and associated torques
change  when varying the relative exchange field
directions between $F_1$ and $F_2$ in an $S F_1 N F_2 S$ junction. 
A supercurrent is generated in the structure by maintaining a phase
difference which we take to be $\Delta\varphi=100^\circ$
in Fig.~\ref{fig13}.
We  rotate in-plane magnetization  $F_1$, while keep that in
$F_2$
fixed along the $y$ (as in Fig.~\ref{fig6}(b)). 
Control of the  free-layer magnetization  can be achieved  experimentally 
via external magnetic fields\cite{ilya}, 
or spin-torque switching\cite{Bauer_nat,brataas_nat}.
In Fig.~\ref{fig13}(a), 
the $x$ component of the local spin current, $S_x$, is shown
throughout the junction as a function of position $X$,
for four values of the $\phi'_1$ angle (we have
$\theta_1=\theta_2=90^\circ$). 
The spin current is again seen to
be a nonconserved quantity within the ferromagnets,
reflecting the existence of a spin-transfer torque.
In the nonmagnetic normal metal connecting the two $F$ regions,
the current is constant, and its value as $\phi'_1$ is varied 
$S_x$ cycles from 
positive  to  negative.
To explore this further,
we examine the total change in spin current
across  $F_1$,
as $\phi'_1$ sweeps from the P to AP state.
This change is related via Eq.~(\ref{sx2}) to the integrated torque
in this region. Hence, in 
Figure \ref{fig13}(b) we plot
$\tau_{x,{\rm tot}}$ 
vs $\phi'_1$ for a wide range of $D_N$.
When the normal metal is absent ($D_N=0$),
the magnitude of the total torque reaches its peak around 
$\phi'_1=90^\circ$,
indicating that this component
of the torque, which  tends 
to align the magnetic moments  of the two $F$
layers is largest when they are mutually orthogonal.
This makes sense physically. 
The presence of even a thin  normal metal spacer
causes  $\tau_{x,{\rm tot}}$ to become 
much smaller (the $D_N=0$ results are plotted 
after dividing them by five) and nearly 
$\pi$-symmetric, so that
now the orthogonal magnetic configuration
produces negligible net torque
within the $F$ layers.
Increasing $D_N$
reduces the ferromagnetic coupling and hence reduces the magnitude
of the mutual torques,
although
the $\pi$-periodicity is retained.
Finally in Figure \ref{fig13}(c) we plot the local
value of the torque, and find its spatial behavior
to be consistent with that  of $S_x$
as given in Eq.~(\ref{sx1}). 

\begin{figure}
\centerline{\includegraphics[width=8.8cm]{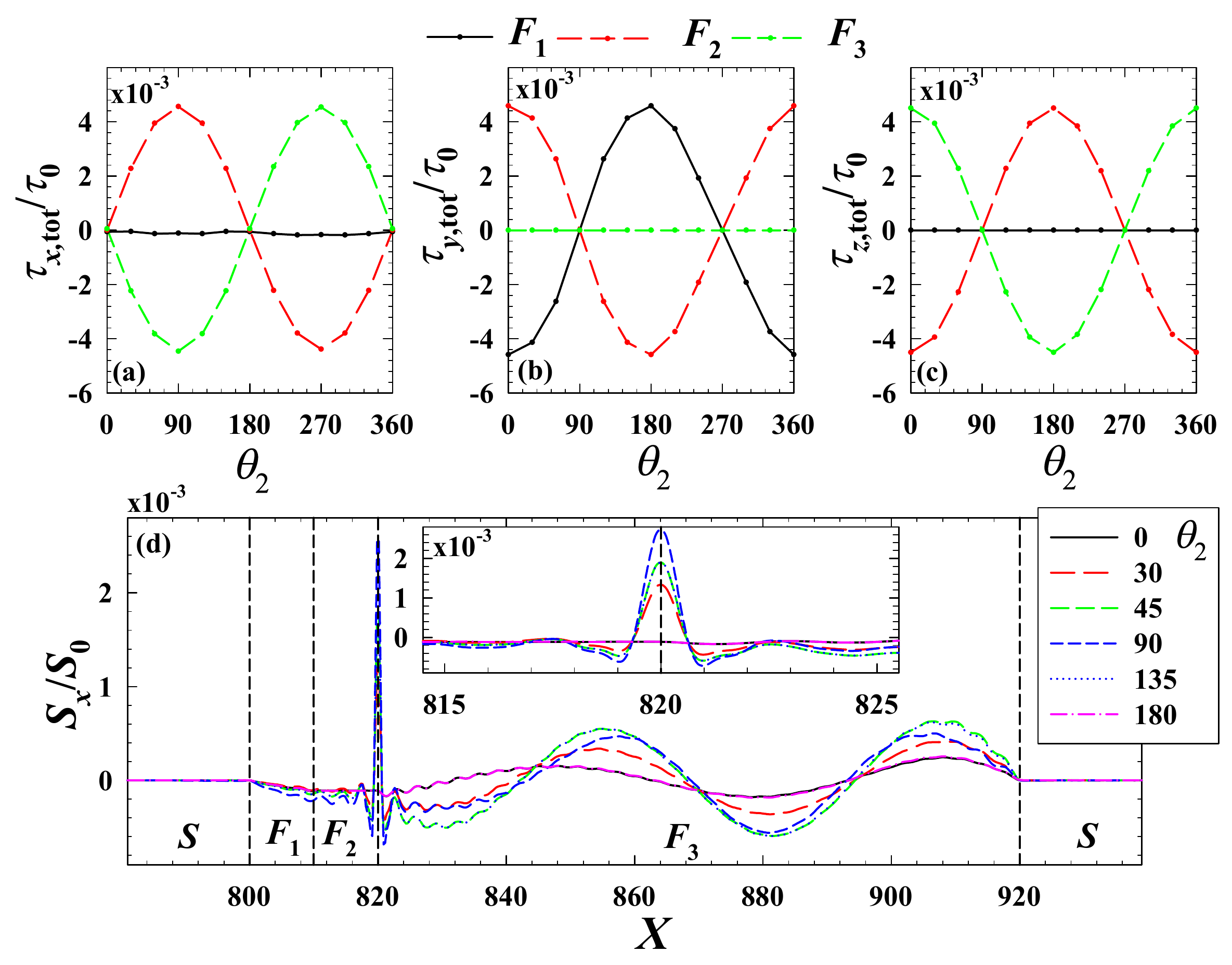}}
\caption{
Top panels: total torque within each of the ferromagnet regions 
(see overhead legend) 
in an $S F_1 F_2 F_3 S$
junction as  the  angle $\theta_2$ varies.
The system parameters are those used in
Fig.~\ref{fig7}(a).
The sum $\tau_{i,{\rm tot}}$ ($i=x,y,x$) over all three 
ferromagnetic regions vanishes for each component.
In the bottom panel, the spatial behavior of the 
normalized $x$-component of the spin current is shown 
throughout the system for a few select magnetization 
orientations $\theta_2$ (see legend). The inset is a magnification of the
region centered around the $F_2/F_3$ interface located at $X=820$. 
Vertical dashed lines in the main plot mark interface locations.
 }   
\label{fig14}
\end{figure}

Finally, we consider the $S F_1 F_2 F_3 S$ system,
studied previously
in Fig.~\ref{fig7}(a), with $D_{F1}=D_{F2}=10$, and $D_{F3}=100$.
A phase difference of $\Delta\varphi=100^\circ$
maintains a constant current throughout the junction,
and there is moderate interface scattering, with $H_B=0.8$.
The magnetization in $F_1$ is along $z$,
and in $F_3$, it is along $y$. The central ferromagnet, $F_2$,
has a magnetization vector that is rotated the $xz$ plane (see Fig.~\ref{fig0}),
so that for $\theta=0^\circ$, it is oriented along $x$, and for $\theta=90^\circ$,
it is aligned along $z$.
For these more complex magnetic configurations,
where  one of the $F$ layers 
possesses an
out-of-plane exchange field,
all three spin components  of the  
current  ${\bm S}$ must be considered. 
The top panels in Fig.~\ref{fig14} 
depict the components of the 
total torque, $\tau_{i,{\rm tot}}$ ($i=x,y,z$)
for each  ferromagnet region, identified in the legend
above these panels.
Since the total torque in a given direction equals (see Eq.~(\ref{sx2})) 
the overall change in spin current, 
and, as explained above 
there is no spin current in the $S$ regions at fixed phase, 
the sum of each  component $\tau_{i,{\rm tot}}$ over
 all $F$ regions must be zero. 
This is seen in these three panels, 
where the oscillatory curves exactly cancel one another.
For each of the three components, we also observe that the total torque 
in either $F_1$ or $F_3$ nearly 
vanishes over the whole angular range of $\theta_2$. 
This follows from the expression for the local torque, Eq.~(\ref{stt}),
which implies  that
$\bm \tau$ is orthogonal to the exchange field vector $\bm h$,
and the magnetization $\bm m$.
For example, considering the leftmost panel 
the only component of the exchange field in $F_1$  is
along $z$, and  
since $\tau_x\sim m_y h_z$,  a $y$ component
of the magnetization is needed in $F_1$
to generate a torque along $x$. 
However
${\bm h}$ in the adjacent $F_2$ rotates solely in the $xz$ plane, and 
thus $m_y$ vanishes in $F_1$ except in a narrow
region near the interface (see Ref.~\onlinecite{wvh14}) resulting in 
a very small value for the averaged $\tau_{x,tot}$.

As mentioned  previously the spin current ${\bm S}$
is a local quantity, and  when it is spatially nonuniform
the resulting torque  influences
the magnetization configurations.  
It is therefore insightful to 
examine also in this case, as we did in Figs.~\ref{fig12}
and \ref{fig13}, 
the spatial behavior of the spin current.
The results are  displayed in
 the bottom panel of Fig.~\ref{fig14}.
 For clarity, we present only the $x$-component, $S_x$,
 since the other spin components behave similarly. 
A few representative angle orientations, $\theta_2$, are considered
(see legend).
As expected, we see that $S_x$ vanishes in the outer $S$ electrodes.
Within the larger $F_3$ region, the spin current undergoes regular oscillations
which are much harder
to distinguish in the  narrow $F_2$ and $F_1$ regions.
To compare with
the previous results,  
we see from this panel that  
the change in the $x$-component of the spin current, $\Delta S_x$ across
the $F_1$ boundaries is negligible, in agreement with 
the results in the left top row panel. 
In $F_2$, this component of the
current is very small near the left interface,
but it increases near the right edge, so that
$\Delta S_x$ agrees well with the $\tau_{x,{\rm tot}}$
 variations in $F_2$ observed in the left top panel.
Also in agreement is the enhancement of the spin current $S_x$ that is
narrowly peaked
at the $F_2/F_3$ interface for $\theta_2$ 
that are near normal to
the plane ($\theta_2\approx 90^\circ$). 
 Finally, since  the interface adjoining 
 $F_3$ and the right $S$ electrode 
 has $S_x=0$,
 the change $\Delta S_x$ in the $F_3$ region is due entirely from the value of 
 the spin current at the $F_2/F_3$ interface, thus resulting in
 $\Delta S_x$  that is exactly opposite to that in $F_2$.

\section{Conclusions}
\label{conclusions}

We have presented here an extensive study of the Josephson
currents flowing in generic ballistic
structures of the $SFS$ type
where the
$F$ regions contain two or three ferromagnetic layers, 
possibly separated by normal spacers.
For the $SFFS$-type spin valves, we study their transport properties 
by considering different in-plane relative magnetization angles.
When the third $F$ layer is present ($SFFFS$), we 
allow the central $F$ region to have out-of-plane magnetic orientation 
while those for the two outer $F$ layers are still in-plane.
Our self consistent formalism ensures\cite{wvh14} that the  charge conservation
law is satisfied and that the 
proper relations that balance the spin transfer torques and 
the gradients of the spin current components hold. Results are
given for a wide range of values of the geometrical and 
orientation parameters, as well as interfacial scattering.

We have organized our results in several subsections. We first
have considered (Sec.~\ref{cpr}) the current-phase relations (CPRs) as 
a function of geometrical parameters (layer thicknesses) at fixed
relative angles between the in-plane magnetizations: parallel (P),  
antiparallel (AP), and normal (N). We find that in general, 
larger geometric asymmetry (the aspect ratio for the thicknesses of 
the two outer $F$ layers)
leads to larger superharmonic
($\pi$ periodic) behavior. This is particularly pronounced in the 
two-magnet case. It is found that the strength of interfacial 
scattering can affect the magnitude of the critical current. 
Next, in Sec.~\ref{mocpr}, we consider the effect
of magnetization misorientation on the CPR. We find that these effects
are profound. 
In particular, by sweeping the relative in-plane angle from P to AP
at fixed phase differences
$\Delta\varphi$ between two $S$ electrodes, the supercurrent flow first 
vanishes at N configurations, followed by reversal of its direction. 
These can be understood to a very large extent by noting 
that the generation of induced spin triplets (studied in Sec.~\ref{itp})
is correlated with magnetic inhomogeneity and, via this phenomenon, to
the CPR relationships.
In Sec.~\ref{itp}, we also presented the local spatial behavior of both the $m=0$ 
and $m=\pm 1$ triplet correlations and carefully quantify each component as
functions of magnetic orientations and $\Delta\varphi$. The results clearly
demonstrate the existence of the singlet-to-triplet conversion in the 
Josephson junctions.
Finally, in Sec.~\ref{st} we have discussed spin
transport and shown that, due to the interaction between the 
charge current and the magnetizations,
both the spin current and the spin transfer
torque (STT) oscillate in the $F$ regions. By varing 
$\Delta\varphi$ or the magnetic misalignment angles, 
the phase of their oscillations can change accordingly. 
In addition, we have shown 
how the spin current gradient equals the spin transfer 
torque. 

We hope that our rather comprehensive study of transport in these 
multilayer structures will guide the experimentalist in choosing optimal 
configurations for building devices such as low dissipation memory
storage units, which are expected to rely on the  behavior of the
Josephson junctions studied here, in particular on the behavior of the
CPR as orientation angles are changed.

\acknowledgments
K.H. is supported in part by ONR and a grant of HPC
resources from the DOD HPCMP. 
K.H. would like to thank M. Alidoust for helpful
discussions.

\appendix

\section{Numerical Procedure}
\label{appB}

Here we discuss some
technical aspects of the numerical procedure used
in calculating the spin and charge currents governed 
by the Andreev bound states.
We first expand\cite{klaus} the quasiparticle amplitudes 
in terms of a complete set:
\begin{align}
\psi_{n }(x) = \sqrt{\frac{2}{d}}\sum_{q=0}^{N} \sin ({k_q x}) \hat{\psi}_{q} (k_q),
\label{FT}
\end{align}
where we use the shorthand notation
$\psi_{n} (x)=(u_{n\uparrow}(x),u_{n\downarrow}(x),v_{n\uparrow}(x),v_{n\downarrow}(x))$,
and $\hat{\psi}_{q} =(\hat{u}_{q\uparrow},\hat{u}_{q\downarrow},\hat{v}_{q\uparrow},\hat{v}_{q\downarrow})$.
We write the wavevector index as, $k_q = q \pi/d$,
so that
$\Delta k_q \equiv k_{q+1}-k_q= \pi/d$. 
Thus
$N$ grid points subdivide the system of width $d$.
We 
take $d$ to be large enough so
that  the results become independent of $d$.
The next step involves Fourier transforming the real-space BdG equations (Eq.~\ref{bogo}),
resulting in the following set of coupled equations in momentum space:
\begin{align}
\label{kbogo}
\begin{pmatrix} 
\hat{H}_0 -\hat{h}_z&-\hat{h}_x+i\hat{h}_y&0&\hat{\Delta} \\
-\hat{h}_x-i\hat{h}_y&\hat{H}_0 +h_z&\hat{\Delta}&0 \\
0&{\hat\Delta}^*&-(\hat{H}_0 -\hat{h}_z)&-\hat{h}_x-i\hat{h}_y \\
{\hat \Delta}^*&0&-\hat{h}_x+i\hat{h}_y&-(\hat{H}_0+\hat{h}_z) \\
\end{pmatrix}
\begin{pmatrix}
\hat{u}_{\uparrow}\\\hat{u}_{\downarrow}\\\hat{v}_{\uparrow}\\\hat{v}_{\downarrow}
\end{pmatrix}
=\epsilon_n
\begin{pmatrix}
\hat{u}_{\uparrow}\\\hat{u}_{\downarrow}\\\hat{v}_{\uparrow}\\\hat{v}_{\downarrow}
\end{pmatrix}.
\end{align}
Here we have defined $\hat{u}_\sigma=(\hat{u}_{1\sigma},\hat{u}_{2\sigma},\ldots, \hat{u}_{N\sigma})$,
$\hat{v}_\sigma=(\hat{v}_{1\sigma},\hat{v}_{2\sigma},\ldots, \hat{v}_{N\sigma})$,
and the matrix elements,
\begin{align}
\label{hok}
\hat{H}_0(q,q')&=\frac{2}{d} \int_{0}^d dx \left(\frac{k_q^2}{2m}+\epsilon_\perp-\mu\right) \sin(k_q x) \sin(k_{q'} x),  \\
\label{delk}
\hat{\Delta}({q,q'}) &= \frac{2}{d}\int_{0}^d dx \Delta(x) \sin(k_q x) \sin(k_{q'} x),  \\
\label{hk}
\hat{h}_{i}({q,q'}) &=\frac{2}{d} \int_{0}^d dx\, h_{i}(x) \sin(k_q x) \sin(k_{q'} x), \quad i=x,y,z,
\end{align}
where $\epsilon_\perp$ is the kinetic energy on the $y-z$ plane. 
Our numerical procedure for calculating the supercurrent involves
initially assuming a constant amplitude
form for the pair potential
in each $S$ layer, but with a total phase
difference $\Delta\varphi$ ($0, \Delta\varphi$ at each $S$ 
region). We then expand the pair potential via Eq.~(\ref{delk}).
Similarly the exchange field and free particle Hamiltonian are
expanded using Eq.~(\ref{hk}) and Eq.~(\ref{hok}) respectively.
We then find the quasiparticle energies and amplitudes 
by
diagonalizing the resultant momentum-space matrix (Eq.~(\ref{kbogo})).
Once the momentum-space wavefunctions and energies are found,
they are transformed back into real-space via Eq.~(\ref{FT}). 
From them,
a new pair potential $\Delta(x)$ is self-consistently determined 
via Eq.~(\ref{del2}) through the entire region except
for a small region (three coherence lengths from the sample 
edges) where the pair potential is fixed to its bulk
absolute value, with phases $0, \Delta\varphi$. 
The newly calculated  $\Delta(x)$ is then used  in the BdG equations 
and the above  
process is repeated iteratively until convergence is achieved.
When determining  
the current phase relations,
$\Delta\varphi$ 
is defined  as the difference in phases between the 
superconductors in the outermost self-consistent regions.
As self-consistency evolves with each iteration,  the
final $\Delta\varphi$ often 
differs slightly 
from the fixed  difference  $\Delta\varphi$ that 
 is set in the non self-consistent regions.
Thus, 
to have $\Delta\varphi$ fixed to a prescribed value 
while varying other parameters, e.g., $\theta_2$, 
additional calculations are needed with slightly different 
initial choices for the phase
$\Delta\varphi$. 
Following the discussion in the main text,
when current is flowing through the junction, the self-consistently 
calculated regions
are always been found to possess the necessary spatially constant current.
The fixed-phase non-self consistent edge regions, 
provide  the physically necessary  source or sink
of current, via the applied electrodes, thus
acting as an effective boundary condition.

\section{Spin Rotation Matrices}
\label{appA}
Here we show how to perform the spin rotations 
for the two triplet components $f_0$ and
$f_1$ (the singlet amplitude is 
of course invariant under spin rotations).
The problem simplifies if all one wishes
is to align the spin quantization axis
with the local magnetization direction: this
affords easier physical interpretation of the results.
The central quantity that we use to perform the desired rotations
is the spin transformation matrix $\mathcal{T}$ in particle-hole
space. 
The quasiparticle amplitudes  transform as,
\begin{align}
\Psi^\prime_n (x)
 = \mathcal{T}
\Psi_n (x). \label{transform}
\end{align}
In our notation
the matrix $\mathcal{T}$  can be written as:
\begin{align}\label{tmatsmall}
\mathcal{T}= &\left[
  \begin{array}{cccc}
   \mathcal{A}& 0  \\
   0 & \mathcal{B}
  \end{array}
\right],
\end{align}
where the submatrices $\mathcal{A}$ and $\mathcal{B}$ are
trigonometric functions solely of the angles that describe the local
magnetization orientation. Expressing the orientation
of the exchange fields in the
regions $F_1$ and $F_2$ in terms of the angles 
 $\theta_i$ and $\phi_i$ introduced in Eq.~(\ref{fields})
we can write 
$\mathcal{A}$ and $\mathcal{B}$ as the following $2\times2$ matrices:
\begin{widetext}
\begin{align}\label{tmat}
\mathcal{A}= &\left[
  \begin{array}{cccc}
   \cos\left({\phi_i}/{2}\right)\sin\left(\theta^+_{i}\right) +i\sin\left({\phi_i}/{2}\right)\sin(\theta^-_i) &
    -\cos({\phi_i}/{2})\sin(\theta^-_i) -i\sin({\phi_i}/{2})\sin(\theta^+_i) \\  \\
    \cos({\phi_i}/{2})\sin(\theta^-_i)-i\sin({\phi_i}/{2})\sin(\theta^+_i)  &
        \cos({\phi_i}/{2})\sin(\theta^+_i)-i\sin({\phi_i}/{2})\sin(\theta^-_i)
  \end{array}
\right],
\end{align}
\begin{align}\label{tmat2}
\mathcal{B}= &\left[
  \begin{array}{cccc}
     \cos\left({\phi_i}/{2}\right)\sin\left(\theta^+_{i}\right) -i\sin\left({\phi_i}/{2}\right)\sin(\theta^-_i) &
   \cos({\phi_i}/{2})\sin(\theta^-_i) -i\sin({\phi_i}/{2})\sin(\theta^+_i) \\  \\
  -\cos({\phi_i}/{2})\sin(\theta^-_i)-i\sin({\phi_i}/{2})\sin(\theta^+_i)  &
    \cos({\phi_i}/{2})\sin(\theta^+_i)+i\sin({\phi_i}/{2})\sin(\theta^-_i)
  \end{array}
\right].
\end{align}
\end{widetext}
Here we have defined $\theta_i^\pm \equiv \theta_i/2\pm\pi/4$.
Using the spin rotation matrix $\mathcal{T}$, we can transform the
original BdG equations ${\cal H}\Psi_n=\epsilon_n\Psi_n$ (Eq.~(\ref{bogo})) by performing the
unitary transformation: ${\cal H}' = \mathcal{T} {\cal H}
\mathcal{T}^{-1}$ (of course we have $\mathcal{T}^\dagger \mathcal{T} =1$). We
then end up with the magnetization effectively along the new $z$ axis
and:
\begin{align}\label{bdg2}
{\cal H}'=
&\left(
  \begin{array}{cccc}
    {\cal H}_0-h &0 & 0 &\Delta  \\
   0& {\cal H}_0+h & \Delta & 0  \\
     0 &\Delta^*& -{\cal H}_0+h & 0\\
    \Delta^* & 0 &0 &  -{\cal H}_0-h
  \end{array}
\right).
\end{align}

One of the benefits of working in this rotated coordinate 
system is that now
the Hamiltonian matrix can be reduced to a smaller $2\times 2$ size by
using symmetry properties that now exist between the quasiparticle amplitudes
and energies.\cite{klaus}
As is the case under all unitary transformations,
the eigenvalues here
are preserved, but the eigenvectors are modified in general
according to Eq.~(\ref{transform}).
Thus for example,
operating on the wavefunctions using Eq.~(\ref{transform}),
and examining the terms involved in calculating the singlet pair
correlations (Eq.~(\ref{del2})),
it is easily shown that for a given set of quantum numbers $n$ and
position $x$, the following relation  between the 
transformed (primed) 
and untransformed quantities holds: $ {u'}_{n\uparrow} 
{v'}_{n\downarrow}^{*}+{u'}_{n\downarrow}{v'}_{n\uparrow}^{*} =
u_{n\uparrow} v_{n\downarrow}^* +u_{n\downarrow}
v_{n\uparrow}^*$. Thus, the terms that dictate the
singlet pairing are invariant
for any choice of quantization axis, transforming as scalars under
spin rotations, as they should.

The terms governing the triplet amplitudes on the other hand are
generally not invariant under the spin-rotation. It is illuminating
to see how both the equal-spin and different spin triplet correlations
transform. The relevant particle-hole products in Eq.~(\ref{f0})
that determine $f_0$, upon the spin transformations obey the following
relationships:
\begin{align}
&{u'}_{n\uparrow} {v'}_{n\downarrow}^{*} -{u'}_{n\downarrow}
{v'}_{n\uparrow}^{*} = \cos\theta_i\left(u_{n\uparrow} v_{n\uparrow}^{*}
+u_{n\downarrow} v_{n\downarrow}^{*}\right) \\ \nonumber
&\hspace{-.1cm}+\sin\theta_i\left[ \cos\phi_i \left(u_{n\uparrow} v_{n\downarrow}^{*}-u_{n\downarrow} v_{n\uparrow}^{*}
\right)\hspace{-.08cm}
+\hspace{-.08cm}i\sin\phi_i\left(u_{n\uparrow} v_{n\uparrow}^{*}-u_{n\downarrow} v_{n\downarrow}^{*}
\right)\right]\hspace{-.07cm}.
\end{align}
Similarly the quasiparticle terms in the sum for $f_1$
(Eq.~(\ref{f1})) transform as:
\begin{align}
&{u'}_{n\uparrow} {v'}_{n\uparrow}^{*} + {u'}_{n\downarrow} {v'}_{n\downarrow}^{*}
=
\sin\theta_i\left(u_{n\uparrow} v_{n\uparrow}^*
+u_{n\downarrow} v_{n\downarrow}^{*}\right) \\ \nonumber 
& \hspace{-.1cm} + \cos\theta_i\left[ \cos \phi_i \left(u_{n\downarrow} v_{n\uparrow}^*
-u_{n\uparrow} v_{n\downarrow}^*\right) \hspace{-.07cm}+ \hspace{-.07cm}i\sin \phi_i
\left(u_{n\downarrow} v_{n\downarrow}^*
-u_{n\uparrow} v_{n\uparrow}^*\right)\right] \hspace{-.07cm}.
\end{align}
Thus the triplet amplitudes $f_0$ and $f_1$ in the rotated system are linear combinations of the $f_1$ and $f_0$
in the original unprimed system (and vice versa). It is a simple matter to go from the rotated to the original system
(and vice versa) by the route expressed in Eq.~(\ref{transform}).


\begin{thebibliography}{99}
\bibitem{convention} The Roman ``N'' in this paper denotes the type of magnetic configuration 
and is not to be confused with the Italic ``$N$'' that represents normal metallic layer.
\end{thebibliography}
\end{document}